\documentstyle[aps,prd,psfig]{revtex}

\newcommand{\bm}[1]{{\mbox{\boldmath$#1$}}}

\begin{document}
\author{Sergei M. Kopeikin\thanks{On leave from: ASC FIAN, Leninskii Prospect, 
53, Moscow, 117924, Russia } and Gerhard
Sch\"afer}
\address{Theoretisch-Physikalisches Institut der 
Friedrich-Schiller-Universit\"at
Jena, Max-Wien-Platz 1, 07743 Jena, Germany}
\author{Carl R. Gwinn}
\address{Department of Physics, Broida Hall, University of California 
at Santa Barbara, Santa Barbara, CA 93106}
\author{T. Marshall Eubanks}
\address{US Naval Observatory, 3450 Massachusetts Avenue, Washington,
DC 20392}
\title{Astrometric and Timing Effects of Gravitational Waves from Localized
Sources}
\maketitle
\date{\today}
\tableofcontents
\newpage
\begin{abstract} 
The extremely high precision of current radio interferometric observations
demands a better theoretical treatment of secondary effects in the
propagation of electromagnetic signals in variable gravitational fields.
Such fields include those
of oscillating and precessing stars, stationary or coalescing binary systems,
and colliding galaxies. Especially important is the problem of propagation
of light rays in the field of gravitational waves emitted by a localized source
of gravitational radiation. 
A consistent approach for a complete and exhaustive solution of this problem 
is developed in the present paper in the first post-Minkowskian and quadrupole 
approximation of General Relativity. This approximation is linear with respect 
to the universal gravitational constant $G$ and accounts for the static 
monopole, spin, and time-dependent quadrupole moments of an isolated system. 
We demonstrate for 
the first time that the equations of light propagation in the retarded 
gravitational field of an arbitrary localized source emitting quadrupolar 
gravitational waves can be integrated exactly in closed form.  
The influence 
of the gravitational field under consideration on the light propagation is 
examined not only in the wave zone but also in cases when light passes through 
the intermediate and near zones of the source. We reproduce the 
known results of integration of equations of light rays, both in a 
stationary gravitational field and in the field of plane gravitational 
waves,
establishing the relationship between our new formalism and 
the
simplified approaches of other authors. Explicit analytic expressions for light
deflection and integrated time delay (Shapiro effect) 
are obtained accounting for all possible retardation effects and arbitrary
relative locations of the source of gravitational waves, the 
source of light
rays , and the observer. 
Coordinate dependent terms in the
expressions for observable quantities are singled out and used for 
physically meaningful interpretation of observable quantities. It is shown 
that the ADM and harmonic gauge
conditions can both 
be satisfied simultaneously outside the source of gravitational
waves. Such ADM-harmonic coordinates are extensively used in the present paper.
Their use drastically simplifies the integration of light
propagation equations and the equations for the motion of light source and
observer in the gravitational field of the source of 
gravitational waves, leading 
to the unique interpretation of observable effects. The two limiting 
cases of small
and large values of impact parameter $d$ are elaborated in more detail. 
It is proved that leading order terms for the effect
of light deflection in the case of small impact parameter
depend neither 
on the radiative part ($\sim 1/d$) of the gravitational field
nor on the intermediate ($\sim 1/d^2$) zone terms, 
confirming a previous result 
in the literature. The main effect rather comes
from the near zone ($\sim 1/d^3$) terms. This property of strong 
suppression of 
the influence of gravitational waves on the propagation of light rays makes 
much
more difficult any 
direct detection of gravitational waves by VLBI or pulsar
timing techniques, 
in contrast to previous claims by other authors.
We also present a thorough-going
analytical treatment of time delay and bending of light in 
the case of large
impact parameter. This exploration
essentially extends previous results regarding propagation of
light rays in the field of a 
plane monochromatic gravitational wave. Explicit 
expressions for Shapiro effect and deflection angle are
obtained in terms of the transverse-traceless (TT) part of the space-space 
components of the metric tensor. We also discuss the relevance 
of the developed formalism for
interpretation of radio interferometric and timing observations, as 
well as for 
data processing algorithms 
for future gravitational wave detectors.
\end{abstract}
\newpage

\section{Introduction}
\subsection{Historical Remarks}

Binary systems are well known sources of periodic gravitational waves. 
Indirect proof of the existence of gravitational waves emitted by binary pulsars
was given by Taylor \cite{1}. However, the direct observation of gravitational waves
still remains the unsolved problem of experimental gravitational physics.
The expected spectrum of gravitational waves
extends from $\sim 10^4$~Hz to $10^{-18}$~Hz \cite{2}, \cite{3}.  
Within that range, the
spectrum of periodic waves from known binary systems extends from
about $10^{-3}$~Hz, the frequency of gravitational radiation from a
contact white-dwarf binary \cite{4}, through the $10^{-4}$ to
$10^{-6}$~Hz range of radiation from main-sequence binaries \cite{5},
to the $10^{-7}$ to $10^{-9}$~Hz frequencies
emitted by binary supermassive black holes postulated to lie in
galactic nuclei \cite{6}. The dimensionless strain
of these waves at the Earth, $h$, may be as great as $10^{-21}$ at the
highest frequencies, and as great as $3\times 10^{-15}$ at the lowest
frequencies in this range.

Sazhin \cite{7} first suggested detection of gravitational waves 
from a binary system using timing observations of a pulsar, the line of sight to which 
passes near the binary. It was shown that 
the integrated time delay for 
propagation of the electromagnetic pulse near the binary is proportional to $1/d^2$
where $d$ is the impact parameter of the unperturbed trajectory of the signal. 
More recently,
Sazhin \& Saphonova \cite{8} made estimates of the probability of
observations of this effect, for pulsars in globular clusters, and
showed that the probability can be high, reaching 0.97 for one cluster. We note
however that mathematical technique worked out in these papers allows 
rigorous treatment only of effects of the 
near-zone, quasi-static quadrupolar part of the
gravitational field and is not enough to make any 
conclusion about actual observability of gravitational waves emitted by a
binary system. 

Wahlquist \cite{9} made another approach to the detection of periodic
gravitational waves, based on Doppler tracking of spacecraft traveling
in deep space. His approach is restricted by the plane gravitational wave
approximation developed earlier by Estabrook \& Wahlquist \cite{10}. 
Tinto (\cite{11}, and references therein) made the most recent theoretical 
contribution in this area. The Doppler tracking technique has been
used in space missions, by seeking the characteristic triple
signature, the presence of which would reveal the influence of a
gravitational wave crossing the line of sight from spacecraft to
observer \cite{12}. 

Quite recently, Braginsky {\it et al.} \cite{13}
(see also \cite{14}) have raised the question of using
astrometry as a detector of stochastic gravitational waves.  This idea
has also been investigated by Kaiser \& Jaffe \cite{15} and, in
particular, by Pyne {\it et al.} \cite{16} and Gwinn {\it et al.} \cite{17} 
who
showed that the overall effect is proportional to the strain of metric
perturbation caused by the plane gravitational wave 
and set observational limits on the energy
density of ultra long gravitational waves present in early universe. Montanari
\cite{18} studied polarization perturbations of free electromagnetic radiation in
the field of a plane gravitational wave and found that the effects are
exceedingly small.

Fakir (\cite{19}, and references therein) has suggested the possibility
of using astrometry to detect periodic variations in 
apparent angular separations of appropriate light sources, caused by
gravitational waves emitted from isolated sources of gravitational
radiation. He was not able to develop a self-consistent approach to tackle the
problem with necessary completeness and rigor. For this reason his estimate of 
the effect is erroneous. Another attempt to work out a more consistent approach
to the calculation of the deflection angle in the field of arbitrary source of
gravitational waves has been undertaken by Durrer \cite{20}. However, the 
calculations have been done only for the plane wave approximation and the result
obtained was extrapolated for the case of
the localized source of gravitational waves without justification. For this
reason the deflection angle was overestimated. The same 
misinterpretation of 
the effect can be found in the paper by Labeyrie \cite{21} who studied a 
photometric modulation of background sources by gravitational waves
emitted by fast binary stars. Because of this, the expected detection of the
gravitational waves from the
observations of the radio source GPS QSO 2022+171 suggested by Pogrebenko
{\it et al.} \cite{22} was not based on firm theoretical ground.

Damour \& Esposito-Far\`ese \cite{23} have studied the deflection of light and 
integrated
time delay caused by the time-dependent gravitational field generated by a
localized material source lying close to the line of sight. They explicitly
took into account the full, retarded gravitational field in 
the near, intermediate, and wave zones. 
Contrary to the claims of Fakir \cite{19} and Durrer
\cite{20} and in agreement with Sazhin's \cite{7} 
calculations, they found that the
deflections due to both the wave-zone gravitational wave and the
intermediate-zone retarded fields vanish exactly. The leading total
time-dependent deflection is given only by the quasi-static, near-zone 
quadrupolar piece of the gravitational field.   

In the present paper we work out an even more systematic approach to the problem.
While Damour \& Esposito-Far\`ese \cite{23} considered both the light source and 
the observer to be located at infinity, and performed their calculations in 
terms of the
spacetime Fourier transform, we do not need these assumptions. Our approach is 
much more general and applicable for any location of the source of light and 
observer in space with respect to the source of gravitational radiation. The
integration technique which we use for finding the solution of equations of 
propagation of light rays was partially employed in \cite{24} and does
not require any implementation of the spacetime Fourier transform.

Section 2 of the present paper discusses equations of propagation of
electromagnetic waves in the 
geometric optics approximation. The metric tensor and
coordinate systems involved in our calculations are described in section 3
along with gauge conditions imposed on the metric tensor. The method of 
integration of the equations of motion with emphasis on specific details of
calculations of particular integrals is given in section 4. Exact solution of
the equations of light propagation and the form of relativistic perturbations
of the light trajectory are obtained in section 5. Section 6 is devoted to
derivation of basic observable relativistic effects - the integrated time delay
and the deflection angle. We find the more precise meaning of quite general 
formulae obtained in the previous section by discussing in section 7 several 
limiting cases in the 
relative configuration of the source of light, the observer, and the source of 
gravitational waves. Section 8 contains concluding remarks. Appendix A compares
results of our calculations with those
by Damour \& Esposito-Far\`ese \cite{23} and proves their gauge invariance.
Appendix B gives more details on the derivation of the ADM-harmonic coordinate
system used in the present paper for interpretation of observed relativistic 
effects.

\subsection{Observational Capabilities}

Calculations of the effects of gravitational waves are of most
interest if they indicate that those can be detected with present
techniques, or foreseeable improvements.  Astrometric precision and
accuracy have evolved rapidly in the last decades, and can be expected
to continue to improve.  In principle, the accuracy attainable with a
given instrument is approximately the angular resolution of the
instrument, divided by the signal-to-noise ratio.  In practice, the
time span of the observations and the angular separation of the source
from reference sources critically affect the attainable accuracy.

Very-long baseline interferometry, or VLBI, attains the highest
angular resolution available on an operational basis.  It achieves
angular resolution set by the diffraction limit, of
$\Delta\theta\approx \lambda/B$, where $B$ is the separation of the
interferometer elements (the baseline), and $\lambda$ is the observing
wavelength.  Practical baselines may be about as long as an Earth
radius, $B\sim 6400$~km; a typical observing wavelength is
$\lambda=3$~cm, yielding angular resolution of 1~milliarcsecond.

Observations of a moderately strong ($\sim 1$~Jy) 
extragalactic source, such as a
quasar, can reach signal-to-noise ratio of several hundred in 5 or 10
minutes, offering potential angular accuracy of microarcseconds.
In principle, a day of integration
with the US Very Long Baseline Array (VLBA)
could yield angular accuracy of about 0.1~microarcseconds.

Observations using the largest radiotelescopes can increase the signal-to-noise
ratio by a factor of $\sim 10\times$.
In practice, a host of geodetic and propagation
effects limit the reproducibility of VLBI astrometry. These factors
must either be measured during the observations, or calculated from
models.  At present, atmospheric stability and changes in source
structure limit reproducibility of measured angles between sources to
about 1~milliarcsecond, over periods of months.
Observations of pairs of radio sources, with separations of 
$\sim 0.5^{\circ}$,
can yield angular accuracy of about 50~microarcseconds,
reproducible over periods of years,
when effects of source structure are included \cite{25}.

Astrophysical H$_2$O masers have extremely high flux densities, of up
to $10^6$~Jy at $\lambda=1.3$~cm.  
In principle, a day of observation of masers with the VLBA could 
yield angular accuracy of a few picoarcseconds.
Observations of masers have attained
reproducibility of better than 10~microarcseconds over several months,
between individual maser spots in a Galactic maser cluster, with
separations of a few arcsec \cite{26}.  Astrometric
observations of extragalactic masers have attained accuracies of
better than 1~microarcsecond, for maser spots separated by
less than 1 arcsec \cite{27}.
Atmospheric variations probably dominate the error budget.

Shorter wavelengths offer potentially higher diffraction-limited
angular resolution, but practical obstacles are severe.  Atmospheric
effects present greater phase changes, on shorter time scales; and
photon shot noise becomes a limiting factor for fainter sources and at
shorter wavelengths.  Optical interferometers in space will probably
equal and exceed the accuracy of VLBI.  For example, the Space
Interferometry Mission (SIM), and the proposed 
European mission GAIA seek to attain
angular accuracy of about 1~microarcsecond in several hours of integration 
\cite{28} - \cite{30}.

Astrometric observations to seek effects of gravitational waves could
attain higher accuracy, at least on shorter timescales.  The periods
of the waves, and of the expected deflection, could be 
short enough to avoid some atmospheric and other
propagation effects.  For known binary systems, the wave period, and
perhaps the phase, are known accurately, permitting search
for deflections at this period. 
Such a ``synchronous'' search would eliminate many noise sources,
allow detection of short-period motions with the sensitivity
resulting from long integrations,
and perhaps allow astrometric accuracy to approach 
the signal-to-noise ratio limit.

\section{Equations of propagation of electromagnetic waves}

We assume the approximation of geometric optics, as the
wavelength of electromagnetic waves used for 
astrometric observations is usually 
much smaller than wavelength of gravitational waves emitted by isolated
astronomical systems like binary stars or supernova 
explosions \cite{2}.  
This allows us to use the
relativistic equation of geodesic motion of a massless particle 
(such as a photon) for description of the process of
propagation of electromagnetic signal from the source of light to the observer at
the Earth. We also assume that space-time is asymptotically flat. This
assumption does not hold for cosmological distances. However, if we neglect
all terms depending on the rate of cosmological expansion and make a
rescaling of time and space coordinates with the cosmological scale 
factor $a(t)$, our formalism will be still valid for application in cosmology. 

We denote spatial coordinates by $x^i={\bf x}
=(x^1,x^2,x^3)$ and time
coordinate $x^0=ct$, where $c$ is the speed of light and $t$ is coordinate
time. 
Let the motion of a photon be defined by fixing the mixed 
initial-boundary conditions 
introduced and extensively used by Brumberg \cite{31} 
\begin{equation}
{\bf x}(t_{0})={\bf x}_{0}, \quad\quad\quad\quad 
{\displaystyle {d{\bf x}(-\infty ) \over dt}}%
={\bf k},
\label{1}
\end{equation}
where ${\bf k}^{2}=1$ and the spatial components of
vectors are denoted by bold letters. These conditions define the coordinates 
${\bf x}%
_{0}$ of the photon at the moment of emission $t_{0}$ and its velocity at
the infinite past and the infinite distance from the origin of the spatial 
coordinates (that is, at past null infinity). 
In what follows we put $c=1$ for convenience.

Equation of propagation of photons in a weak gravitational field
is given in
the first post-Minkowskian approximation by the 
formula \cite{31,32}:
\begin{eqnarray}\label{2}
\ddot{x}^{i}(t)&=&\frac{1}{2}g_{00,i}-g_{0i,t}-\frac{1}{2}g_{00,t}\dot{x}^i-
g_{ik,t}\dot{x}^k-\left(g_{0i,k}-g_{0k,i}\right)\dot{x}^k-\\\nonumber 
\\ \nonumber&&\mbox{} 
g_{00,k}\dot{x}^k\dot{x}^i-\left(g_{ik,j}-\frac{1}{2}g_{kj,i}\right)
\dot{x}^k\dot{x}^j+
\left(\frac{1}{2}g_{kj,t}-g_{0k,j}\right)\dot{x}^k\dot{x}^j\dot{x}^i,
\end{eqnarray}
where the $g_{00}$, $g_{0i}$, $g_{ij}$ are components of metric tensor, 
fully determined by
the given distribution and motion of mass inside the source of
gravitational field, dots over vectors denote 
the total derivative with respect to
time, and commas indicate partial derivatives with respect to 
spatial or time coordinates; that is, for any function $f_{,i}=
{\partial}f/{\partial} x^i$, $f_{,t}={\partial}f/{\partial} t$.
Hereafter  repeated latin indices mean summation from $1$
to $3$. The given equation is valid in arbitrary coordinates (gauges) and
represents the ordinary second order differential equation for light
propagation. 

The right-hand side of equation (\ref{2}) includes terms which
depend on the coordinate velocity $\dot{x}^{i}$ of the photon,
in the weak-field approximation approximately equal to the speed of light $c$. 
We restrict ourselves to finding a
solution of equation (\ref{2}) only in the first linear approximation with
respect to the universal gravitational constant $G$. For this reason, when solving
equation (\ref{2}), only one
iteration is enough and it is admissible to make the 
replacement $\dot{x}^{i}=k^{i}$ in
the right-hand side of the equation. The result of this approach is:
\begin{eqnarray}\label{pol}
\ddot{x}^{i}(t)&=&\frac{1}{2}g_{00,i}-g_{0i,t}-\frac{1}{2}g_{00,t}k^i-
g_{ij,t}k^j-\left(g_{0i,j}-g_{0j,i}\right)k^j-\\\nonumber 
\\ \nonumber &&\mbox
g_{00,j}k^j k^i-\left(g_{ip,j}-\frac{1}{2}g_{pj,i}\right)
k^p k^j+
\left(\frac{1}{2}g_{pj,t}-g_{0p,j}\right)k^p k^j k^i\;.
\end{eqnarray}
This equation must be solved to obtain a perturbed trajectory of
the photon propagating through the gravitational field of an isolated
astronomical system emitting gravitational waves. 
To accomplish this task one needs a mathematical expression for the 
metric tensor.

\section{Metric Tensor and Coordinate Systems}

Let us chose the origin of the asymptotically flat coordinate frame at the 
center of mass-energy of the isolated astronomical system and impose the 
de-Donder (harmonic) gauge conditions on components of the
``canonical" metric tensor. We assume that gravitational field is weak and
the metric of spacetime $g_{\alpha\beta}$ is written as a sum of 
the Minkowski metric $\eta_{\alpha\beta}={\rm diag}(-1,1,1,1)$ plus a 
small perturbation $h_{\alpha\beta}$:
\vspace{0.3 cm}
\begin{eqnarray}
\label{rew}
g_{\alpha\beta}&=&\eta_{\alpha\beta}+h_{\alpha\beta},
\end{eqnarray}
where the Greek indices run from 0 to 3.
The most general expression for 
the linearized 
metric tensor, generated by a system emitting gravitational waves, 
in terms of its symmetric and trace-free (STF) mass and spin multipole moments 
is given
by Thorne \cite{33} (see also \cite{34,35}). It can be written as
\vspace{0.3 cm}
\begin{eqnarray}\label{metric}
h_{\alpha\beta}&=&h_{\alpha\beta}^{can.}+\nabla_{\beta}w_{\alpha}+
\nabla_{\alpha}w_{\beta}\;,
\end{eqnarray}
where $\nabla_{\alpha}=\partial/\partial x^{\alpha}$. The ``canonical'' form of 
the metric tensor perturbations in harmonic gauge reads as follows \cite{36}
\vspace{0.3 cm}
\begin{eqnarray}\label{4}
h_{00}^{can.}&=&\frac{2{\cal M}}{r}+2\displaystyle
{\sum_{l=2}^{\infty}}\frac{(-1)^l}{l!}
\left[\frac{{\cal I}_{A_l}(t-r)}{r}\right]_{,A_l}\;,\\
\nonumber\\
h_{0i}^{can.}&=&-\frac{2\epsilon_{ipq}{\cal S}_p N_q}{r^2} -\\\nonumber \\
\nonumber
& &\mbox{}4\displaystyle{\sum_{l=2}^{\infty}}\frac{(-1)^l l}{(l+1)!}
\left[\frac{\epsilon_{ipq}{\cal S}_{pA_{l-1}}(t-r)}{r}\right]_{,qA_{l-1}}+
4\displaystyle{\sum_{l=2}^{\infty}}\frac{(-1)^l }{l!}
\left[\frac{\dot{\cal {I}}_{iA_{l-1}}(t-r)}{r}\right]_{,A_{l-1}}\;,\\
\nonumber\\\label{6}
h_{ij}^{can.}&=&\delta_{ij}\biggl\{\frac{2{\cal M}}{r}+2\displaystyle{\sum_{l=2}^{\infty}}
\frac{(-1)^l}{l!}
\left[\frac{{\cal I}_{A_l}(t-r)}{r}\right]_{,A_l} \biggr\}+ \\ 
\nonumber\\ \nonumber
& &\mbox{} 4\displaystyle{\sum_{l=2}^{\infty}}\frac{(-1)^l }{l!}
\left[\frac{\ddot{\cal {I}}_{ijA_{l-2}}(t-r)}{r}\right]_{,A_{l-2}}-
8\displaystyle{\sum_{l=2}^{\infty}}\frac{l(-1)^l l}{(l+1)!}
\left[\frac{\epsilon_{pq(i}\dot{\cal {S}}_{j)pA_{l-2}}(t-r)}{r}
\right]_{,qA_{l-2}}\;.
\vspace{1 cm}
\end{eqnarray}
where $N^i=x^i/r$, and the round brackets around indices in equation
({\ref{6}) means symmetrization; that is, for any two indices
$T_{(ij)}=\frac{1}{2}(T_{ij}+T_{ji})$.
In the pure harmonic gauge the functions $w^0$, $w^i$ are solutions of the 
homogeneous d'Alembert's equation and are 
given by the expressions
\vspace{0.3 cm}
\begin{eqnarray}\label{poh}
w^0&=&\displaystyle{\sum_{l=0}^{\infty}}
\left[\frac{{\cal W}_{A_l}(t-r)}{r}\right]_{,A_l}\;,\\
\nonumber\\\nonumber\\\label{boh}
w^i&=&\displaystyle{\sum_{l=0}^{\infty}}
\left[\frac{{\cal X}_{A_l}(t-r)}{r}\right]_{,iA_l}+
\displaystyle{\sum_{l=1}^{\infty}}\biggl\{
\left[\frac{{\cal Y}_{iA_{l-1}}(t-r)}{r}\right]_{,A_{l-1}}+
\left[\epsilon_{ipq}\frac{{\cal
Z}_{qA_{l-1}}(t-r)}{r}\right]_{,pA_{l-1}}\biggr\}\;,
\end{eqnarray}
where ${\cal W}_{A_l}$, ${\cal X}_{A_l}$, ${\cal Y}_{iA_{l-1}}$, and ${\cal
Z}_{qA_{l-1}}$ are arbitrary functions of time. Their specific choice will be
made later on in the discussion regarding the interpretation of observable 
effects. 
In equations (\ref{4})-(\ref{boh}), we adopt the notation:
$A_l=a_1a_2...a_l$ is a polyindex, ${\cal M}$ is the total (Tolman or ADM) mass
of the system, ${\cal I}_{A_l}$ and ${\cal{S}}_{A_l}$ are the STF mass and spin
gravitational multipoles, and ${\cal W}_{A_l}$, ${\cal X}_{A_l}$, 
${\cal Y}_{A_l}$, 
${\cal Z}_{A_l}$ are multipoles which reflect the freedom of coordinate
transformations. These multipoles can be eliminated from the metric using
the transformation
\begin{eqnarray}\label{coortr}
x'^{\alpha}=x^{\alpha}-w^{\alpha}\;,
\end{eqnarray}
relating an original harmonic coordinate system $x^{\alpha}$ to 
another harmonic one $x'^{\alpha}$, in 
which only the ``canonical" part of the
metric is present. 

However, we would like to emphasize that, in general, equation (\ref{metric}) 
holds in an arbitrary gauge.
Particular examples of functions $w^0$ and $w^i$ in harmonic gauge are given in
equations (\ref{poh})-(\ref{boh}). Other expressions for $w^0$ and $w^i$ in 
the ADM (Arnowitt-Deser-Misner)
gauge \cite{37} are given in Appendix B wherein we also 
prove that it is possible to
choose functions $w^0$ and $w^i$ in such a way that ADM and harmonic gauge
conditions will be satisfied simultaneously. This means that the 
classes of harmonic
and ADM coordinates overlap. The discussion of different gauges is
helpful for giving a unique interpretation of observable effects by
properly fixing the coordinate degrees of freedom in corresponding
quantities \cite{38}.

The  STF cartesian tensor 
has
a special algebraic structure which eliminates all reducible parts of the
tensor and leaves only the irreducible part having the highest rank
\cite{33,40}. In other words, contraction over of any two indices
of STF tensor gives identically zero. It is worth noting
the absence of
the dipole mass multipole ${\cal{I}}_i$ in equations (\ref{4})-(\ref{6}) 
which is identically 
zero, due to the choice of the origin of coordinate system
at the center of mass of the gravitating system. We also
stress that the multipoles in the linearized metric (\ref{4})-(\ref{boh})
depend on the ``retarded time'' $t-r$. At first sight
this dependence 
seems to make subsequent calculations more difficult. However, just the 
opposite
happens and the dependence of the multipoles on the retarded time makes
the calculations simpler.  

In what follows we consider the concrete case of a localized deflector
emitting gravitational waves. In this section we restrict ourselves to
considering the
influence of gravitational field of the deflector 
on the propagation of electromagnetic signals made by its total constant
mass $M$, spin ${\cal{S}}$, and time-dependent
quadrupole
moment ${\cal{I}}_{ij}(t-r)$ only. This simplifies the expressions 
(\ref{4})-(\ref{6}) for the metric tensor, which are reduced to the 
expressions
\vspace{0.3 cm}
\begin{eqnarray}
h_{00}^{can.}&=&\frac{2{\cal M}}{r}+\nabla_p\nabla_q
\left[\frac{{\cal I}_{pq}(t-r)}{r}\right]\;,
\label{7}\\
\nonumber\\\nonumber\\
h_{0i}^{can.}&=&-\frac{2\epsilon_{ipq}{\cal S}_p N_q}{r^2}+
2\nabla_j\left[\frac{\dot{\cal {I}}_{ij}(t-r)}{r}\right]\;,
\label{8}\\
\nonumber\\\nonumber\\
h_{ij}^{can.}&=&\delta_{ij}h_{00}^{can.}+q_{ij}^{can.}\;,
\label{9}\\
\nonumber
\end{eqnarray}
where\vspace{0.3 cm}
\begin{eqnarray}\label{qij}
q_{ij}^{can.}&=&\frac{2}{r}\ddot{\cal{I}}_{ij}(t-r)\;.\\\nonumber
\end{eqnarray}
Herein terms depending on ${\cal M}$ and ${\cal S}_i$ are static and produce 
well-known effects in the propagation of light rays. Retarded terms that are 
inversely 
proportional to the distance $r$ from the gravitating system describe the pure 
gravitational-wave part of the metric. 

Let us stress that in the harmonic coordinate system the
gravitational-wave part of the metric tensor is present in all of its components
and is expressed through the second time derivative of 
the quadrupole moment \cite{33}.
If we choose the ADM gauge \cite{36} 
it is possible to eliminate the 
gravitational wave terms from the $h_{00}^{can.}$ and $h_{0i}^{can.}$ 
components of the metric
tensor and to bring all of them to $h_{ij}^{adm}$ \cite{41}. 
Then $h_{00}^{adm}$ and $h_{0i}^{adm}$ 
depend only on the ``instantaneous" time $t$ 
and not on the retarded time $t-r$ (see Appendix B). In combining the ADM gauge
with the harmonic gauge an even simpler representation is possible where 
$h_{00}$ and $h_{0i}$ do not depend on time at all. However, the
transformation from the canonical form of metric (\ref{7})-(\ref{qij}) to the
ADM-harmonic form includes integration of the quadrupole moment 
with respect to time.
Appendix B gives a more detailed study of this procedure. 

One might ask whether the ADM or harmonic coordinate system 
is more preferable, for
the adequate physical treatment of the relativistic time delay and 
deflection of light rays in the field of gravitational waves emitted by 
a localized source. Our point of view is that the coordinate system should be
chosen in such a way to be simultaneously both  ADM and harmonic.
The
reason for this
is that an observer who is at rest with respect to the ADM coordinate
system does not feel the gravitational force caused by gravitational waves. 
This
means that if the instantaneous 
gravitational field of the localized source may be
neglected, the observer fixed with respect to the ADM system can be considered
 to be in free fall. Hence, no artificial 
force need be 
applied to the observer in order to
keep him at rest at the fixed coordinate point. The motion 
of such an observer is
described by the extremely simple 
equation $x^i={\rm const}$ and there is no need 
to account for  kinematic effects associated with the observer's motion. 
All these advantages are lost
in the ``canonical" harmonic gauge. An observer fixed with respect to that
coordinate system must be kept at a fixed coordinate 
point by some external 
force to prevent his motion under the influence of gravitational waves. 
The existence of such
a force is unnatural from physical and astronomical points of view. 
On the other hand, the ``canonical" harmonic gauge has the advantage 
of 
a much simpler integration of the equations of light propagation than  the 
``canonical" ADM gauge.
One can see that the``canonical" ADM metric coefficients 
(\ref{a1a})-(\ref{a2a})
contain functions which depend on time $t$ only. As will be clear from the
procedure of integration of equations of light propagation described in the
next section such ``instantaneous" functions of time do not permit explicit
integration of each specific term (only after summing all
terms is the explicit integration possible). 
Fortunately, the classes of ADM and harmonic
coordinate systems overlap and, for this reason, we can substantially benefit
by 
choosing a coordinate system that is simultaneously both ADM and harmonic. This 
allows us to proceed in the following way. First we integrate equations of 
light
propagation in the harmonic gauge and then apply coordinate transformations
(\ref{ttt})-(\ref{kkk}) which transform the 
pure harmonic coordinate system to 
the ADM one without violation of the 
harmonic gauge conditions. This simplifies 
the treatment of observable effects drastically.   

\section{Method of Integration of the Equations of Motion}
\subsection{Useful Relationships}

We introduce astronomical coordinates ${\bf x}\equiv x^i=(x^1,x^2,x^3)$ 
corresponding to the plane of the sky of the 
observer and based on
a triad of the unit vectors $({\bf I}_0,{\bf J}_0,{\bf K}_0)$. The vector ${\bf
K}_0$ points from the observer toward the deflector, and the vectors ${\bf I}_0$
and ${\bf J}_0$ lie in the plane of the sky, being orthogonal
to vector ${\bf K}_0$. The vector ${\bf I}_0$ is directed to the east, and 
${\bf J}_0$ points towards the north celestial pole. The origin of the
coordinate system is chosen to lie 
at the barycenter of the deflector which emits
gravitational waves (see Figure \ref{bundle}). 

Another reference frame based on
a triad of the unit vectors $({\bf I},{\bf J},{\bf K})$ 
rotated with respect to vectors 
$({\bf I}_0,{\bf J}_0,{\bf K}_0)$
is useful as well.
The vector ${\bf K}$ points from the observer toward the source of light, and
the vectors ${\bf I}$ and ${\bf J}$ lie in the plane of the sky, 
being orthogonal
to vector ${\bf K}$, which is different from the plane of the sky being
orthogonal to vector ${\bf K}_0$. This is because the ``plane of the sky" is
actually a sphere, and vectors ${\bf K}$ and ${\bf K}_0$ point in different
directions.
Mutual orientation of one triad with respect to another one is
determined by the following equations
\begin{eqnarray}
\label{rotation}
{\bf I}_0&=&\hspace{0.3 cm}{\bf I}\cos\Omega+{\bf J}\sin\Omega\;,\\
{\bf J}_0&=&-{\bf I}\cos\theta\sin\Omega+{\bf J}\cos\theta\cos\Omega
+{\bf K}\sin\theta\;,\\
{\bf K}_0&=&\hspace{0.3 cm}{\bf I}\sin\theta\sin\Omega-{\bf J}\sin\theta\cos\Omega+{\bf
K}\cos\theta\;,
\end{eqnarray}
where rotational angles $\Omega$ and $\theta$ are constant.  

To integrate the equations of propagation of electromagnetic waves in curved
space-time we must resort to an approximation method. In the Newtonian
approximation, the unperturbed trajectory of the light ray is a straight line:
\begin{equation}
x^i(t)=x^i_N(t)=x^i_0+k^i\left(t-t_0\right),
\label{15}
\end{equation}
where $t_0$ is the instant of time of the photon emission from the point with 
spatial
coordinates $x^i_0$, and $k^i={\bf k}$ is a constant unit vector tangent
to the unperturbed trajectory and directed from the point of emission to the
point of observation of photon (the vector ${\bf k}\approx -{\bf K}$). 
In the Newtonian approximation, the coordinate
speed of the photon $\dot{x}^i=k^i$ and is considered to be constant.

It is convenient to introduce a new independent parameter $\tau$ along the
photon's trajectory according to the rule \cite{24}
\begin{equation}
\label{16}
\tau \equiv {\bf k}\cdot{\bf x}=t-t_0+{\bf k}\cdot{\bf x}_0,
\end{equation}
where the dot symbol between two vectors denotes the Euclidean dot product of
two vectors. The moment $t_0$ of the signal's emission corresponds to the
numerical value of the parameter $\tau_0={\bf k}\cdot{\bf x}_0$, and
the moment $t^{\ast}$ 
of the closest approach of the
unperturbed trajectory of the photon to the origin of the coordinate system
corresponds to the value $\tau=0$
(note that $\tau_0 < 0$ if the source of light is behind the localaized source
of gravitational waves). Thus, we find
\vspace{0.3 cm}
\begin{equation}
\label{mom}
\tau=t-t^{\ast},\hspace{2 cm}\tau_0=t_0-t^{\ast}.
\end{equation}

The variable $\tau$ is negative from the point of emission up to the point of the
closest approach, and is positive otherwise. 
The differential identity $dt= d\tau$ is valid and for this reason 
the integration along ray's path
with respect to time $t$ can be replaced by the integration with respect to
parameter $\tau$.
Using parameter $\tau$, the equation of the unperturbed trajectory of light ray can
be represented as
\begin{equation}
x^i(\tau)=x^i_N(\tau)=k^i \tau+\xi^i,
\label{17a}
\end{equation}
and the distance, $r$, of the photon from the origin of coordinate system is
given by
\begin{equation}
\label{17}
r=r_N(\tau)=\sqrt{\tau^2+d^2},
\end{equation}
where the length of the constant (for a chosen light ray) 
transverse vector ${\bm{\xi}}={\bf k}\times ({\bf x}_0
\times {\bf k})={\bf k}\times ({\bf x}\times {\bf k})$ 
is called the impact parameter of the unperturbed trajectrory of
the light ray, 
$d=|{\bm{\xi}}|$, and the symbol $``\times"$ between two
vectors denotes the Euclidean cross product. It is worth
emphasizing that the vector $\xi^i$ is directed from the origin of the coordinate
system toward the point of the closest approach of the unperturbed path of 
light ray to that origin.
The relations
\begin{equation}
\label{19}
r+\tau=\frac{d^2}{r-\tau},\hspace{2 cm}r_0+\tau_0=\frac{d^2}{r_0-\tau_0},
\end{equation}
also hold, and they are useful for presenting the 
results of integration of the light ray
equations in different form. In particular, if we
assume the strong inequalities  $d\ll r$, and $d\ll r_0$ 
to hold, then
\begin{equation}
\label{19a}
\tau=r-\frac{d^2}{2r}+...,\hspace{2 cm}\tau_0=-r_0+\frac{d^2}{2r_0}+...,
\end{equation}
which clearly shows that at the moment of light reception $\tau$ 
is positive and at that of light emission $\tau_0$ is negative.

Let us consider a set of curves $x^i(\tau)=k^i\tau+\xi^i$ with different 
values of
vectors $k^i$ and $\xi^i$. The vector field $k^i$, defined along the curve
$x^i(\tau)$, describes the direction of a bundle of light rays along the curve,
and introduces a natural ``2+1" 
splitting of 3-dimensional space. The vector $\xi^i$, on the plane 
orthogonal to the bundle of light rays, is a point of intersection
of any of those rays with that plane (see Figure \ref{bundle}). 
This vector does not depend on $\tau$ and 
can be defined, as in equation (\ref{17a}),
by the relationship
\vspace{0.3 cm}
\begin{eqnarray}
\label{addi}
\xi^i&=&P^i_{\mbox{}j} x^j\hspace {0.5 cm},
\end{eqnarray}
\vspace{0.3 cm}
where
\vspace{0.3 cm}
\begin{equation}
P_{ij}=\delta_{ij}-k_{i}k_{j}\hspace{0.5 cm},   
\label{19aa}
\end{equation}
is the projection operator onto the plane  orthogonal to the vector $%
k^{i}$.  The operator has only two algebraically independent components 
and satisfies the
relationship
\vspace{0.3 cm}
\begin{eqnarray}
\label{ghu}
P^i_{\mbox{}k} P^k_{\mbox{}j}&=&P^i_{\mbox{}j}\hspace{0.5 cm}.
\end{eqnarray}
Because of this property we can recast equation (\ref{addi}) into the
form
\vspace{0.3 cm}
\begin{eqnarray}
\xi^i&=&P^i_{\mbox{}j} \xi^j\hspace{0.5 cm},
\end{eqnarray} 
which shows explicitly that the vector $\xi^i$ is constrained to lie in a
2-dimensional plane.
Thus, we immediately have for the operation of partial differentiation in this plane
\begin{equation}
\label{18}
\frac{\partial\xi^i}{\partial\xi^j}=P_j^i=P^{ij}=P_{ij}.
\end{equation}
It is worth noting that the projection operator can be used  to raise and
lower indices of any geometrical object lying in the plane  orthogonal to
vector $k^i$.

In what follows, it is convenient to consider the spatial components 
of coordinates $\xi^i$ as formally independent with subsequent projection onto 
the plane when doing differentiation with respect to $\xi^i$. Therefore 
we always use the operator of differentiation with respect to $\xi^i$ in
combination with the projection operator $P^i_j$. For example, before the
projection we treat 
\begin{equation}\nonumber
\frac{\partial\xi^i}{\partial\xi^j}=\delta^i_j\;,
\end{equation}
and for the same expression with subsequent projection
\begin{equation}
\nonumber
P^q_j\frac{\partial\xi^i}{\partial\xi^q}=P^i_j\;,
\end{equation}
which agrees with equations (\ref{ghu}) and (\ref{18}).
Moreover, the following rule of differentiation
for an arbitrary smooth function $F(t,{\bf x})$ holds
\begin{equation}
\label{20}
\left[\left(\frac{\partial}{\partial x^i}+k_i\frac{\partial}{\partial t}\right)
F\left(t,{\bf x}\right)\right]_{{\bf x}={\bf x}_0+{\bf k}(t-t_0)}=
\left(P^j_i\frac{\partial}{\partial \xi^j}+k_i\frac{\partial}{\partial \tau}\right)
F\left[\tau,{\bm{\xi}}+{\bf k}\tau\right],
\end{equation}
Equation (\ref{20}) is a generalization of the corresponding
formula 
introduced by Kopeikin (\cite{24}, equation (20)) for
functions which do not depend explicitly on time $t$. It is worth noting that
in the left-hand side of formula (\ref{20}) one has first to 
differentiate the function $F(t,{\bf x})$
with respect to time $t$ and spatial coordinates $x^i$ and, then, to make 
the substitution 
${\bf x}={\bf x}_0+{\bf k}(t-t_0)$. However, one makes 
corresponding substitutions in the right-hand side of the formula (\ref{20})
first and only afterwards takes derivatives.

It is useful to stress again that because the coordinates $\xi^i$ lie 
in the plane 
orthogonal to the vector $k^i$
only two of the three
$\xi^1,\xi^2,\xi^3$ are, in fact, independent. We also stress that the
new variables $\xi^i$ and $\tau$ are independent as well. For this reason,
the integration of any function, which can be represented as a  
time derivative with respect to the parameter $\tau$, is always 
quite straightforward
\begin{equation}
\label{22}
{\int}\frac{\partial}{\partial \tau}F(\tau,{\bm{\xi}})d\tau=
F(\tau,{\bm{\xi}})+C({\bm{\xi}}),
\end{equation}
where $C({\bm{\xi}})$ is an arbitrary function of the constant impact
parameter. Moreover, as the vector $\xi^i$ does not depend on time $\tau$, 
the partial derivatives with respect to $\xi^i$ can be
removed from within the time integrals when calculating them along the photon's
trajectory, that is 
\begin{equation}
\label{22aa}
{\int}\frac{\partial}{\partial\xi^i}F(\tau,{\bm{\xi}})d\tau=
\frac{\partial}{\partial\xi^i}{\int}F(\tau,{\bm{\xi}})d\tau.
\end{equation}
Because of these advantages the new independent 
coordinates
$\tau$ and $\xi^i$ are quite useful in calculations. The usefulness of the variables  
$\tau$ and $\xi^i$ has been also recognized by 
Damour \& Esposito-Far\`{e}se \cite{23}.

The equations of motion of light rays (\ref{pol}) in terms of parameters ${\bm
\xi}$ and $\tau$ are simpler, and after accounting for a freedom in gauge
transformations and implementation of relationship
(\ref{20}) assume the 
form \cite{42}
\begin{eqnarray}
\label{eqnm}
\ddot{x}^{i}(\tau)&=&\frac{1}{2}{\hat{\partial}}_i h_{00}^{can.}
-{\hat{\partial}}_{\tau}h_{0i}^{can.}-
\frac{1}{2}k^i {\hat{\partial}}_{\tau}h_{00}^{can.}-k^j
{\hat{\partial}}_{\tau}h_{ij}^{can.}+
k^j{\hat{\partial}}_{i}h_{0j}^{can.}+\\\nonumber\mbox{}&&
\frac{1}{2}\left({\hat{\partial}}_i+k^i{\hat{\partial}}_{\tau}\right)
k^p k^q h_{pq}^{can.}
-{\hat{\partial}}_{\tau\tau}\left(w^i\;-\;k^i\; w^0\right)\;,\\\nonumber
\end{eqnarray} 
where the following notations are used:
$\hat{\partial}_i \equiv P_{ij}\partial/\partial\xi^j\;$, 
$\hat{\partial}_{\tau} \equiv 
\partial/\partial\tau$. Let us
emphasize once again that the representation of equation (\ref{eqnm}) is valid in 
an arbitrary coordinate system
and all metric coefficients are taken 
along 
the unperturbed trajectory of propagation of the light ray; 
that is, 
$h_{\alpha\beta}(t,{\bf x})=h_{\alpha\beta}(\tau, {\bm {\xi}}+{\bf k}\tau)$.
We also remark that the right-hand side of equation (\ref{eqnm}) contains 
only spatial partial
derivatives with the same index ``$i$'' as does 
the left-hand side of the equation.
This contrasts with equation (\ref{pol}) where the 
indices of spatial derivatives are
mixed. 
Equation (\ref{eqnm}) will be used in sections 5 and 6 
for a general
treatment of gravitational perturbations of the photon's trajectory 
and discussion of relativistic time delay and angle of light deflection.

Another useful form of equation (\ref{eqnm}) may be obtained if one introduces
the four-vector $k^{\alpha}=(1,k^i)$. Then we find
\begin{eqnarray}
\label{neweq}
\ddot{x}^{i}(\tau)&=&\frac{1}{2}k^{\alpha}k^{\beta}
{\hat{\partial}}_i h_{\alpha\beta}^{can.}-{\hat{\partial}}_{\tau}\left(
k^{\alpha}h_{i\alpha}^{can.}-\frac{1}{2}k^ik^j k^p q_{jp}^{can.}\right)
-{\hat{\partial}}_{\tau\tau}\left(w^i\;-\;k^i\; w^0\right)\;.
\end{eqnarray}
This form of the 
equation clearly shows that only the first term on the right-hand
side contributes to the deflection of light, 
if observer and source of light are
at infinity. Indeed, one integration of (\ref{neweq}) with respect to time
from $-\infty$ to $+\infty$ brings all  first and second time
derivatives to zero, 
due to the asymptotic flatness of the metric tensor. This makes
a connection between the formalism of the present paper and that 
of Damour \& Esposito-Far\`ese \cite{23} (see also Appendix A).

\subsection{Calculation of Integrals from the Static Part of the Gravitational
Field}

The static part of the  gravitational field of the deflector contributes
to perturbations of light's ray
trajectory, defined by 
the 
following 
indefinite integrals \cite{24}
\begin{equation}
\label{31}
A(\tau,{\bm{\xi}})\equiv{\int}\frac{d\tau}{r}
={\int}
\frac{d\tau}{\sqrt{d^2+\tau^2}}=-\ln\left(%
\sqrt{d^2+\tau^2}-\tau\right),
\end{equation}
\begin{equation}
\label{32}
B(\tau,{\bm{\xi}})\equiv{\int}A(\tau,{\bm{\xi}})d\tau=
-\tau \ln\left(\sqrt{d^2+\tau^2}-\tau\right)-\sqrt{d^2+\tau^2},
\end{equation}
where we have omitted constants of integration which are absorbed by
re-definition of constants of integration of unperturbed light trajectory
(\ref{15}). 
Integrals (\ref{31}), (\ref{32}) are formally divergent at the lower limit.
However, this divergence is not dangerous 
for setting the second of the boundary conditions (\ref{1})
because 
only derivatives of the integral (\ref{31}) appear
in the result of the first time integration of the equations of motion of 
light rays,
eliminating the divergent part of the
integral \cite{43}. With this in mind,
it is easy to prove that integrals (\ref{31}), (\ref{32}) are in agreement with 
the boundary conditions (\ref{1}).

\subsection{Calculation of Integrals from Time Dependent Part of Gravitational
Field}

One meets two ways of calculation of integrals in finding the path
of 
propagation of light in the gravitational field of 
a localized source emitting
gravitational waves. The first method relies upon the use of the Fourier transform
(\ref{fur}) and allows one, at least in principle, to calculate all integrals
explicitly
if one knows the specific structure of the Fourier image of the
quadrupole moment of the deflector \cite{44}. The advantage of the second method is based on
the fact that one deals with the metric depending on retarded time only. 
This allows one to make  a special
transformation of variables within the integral
which excludes any dependence of the 
integrands on the
impact parameter, 
and transfers it to the limits of the integrals. Thus, partial
derivatives of the integrals can be calculated explicitly without
assumptions about the 
structure of the quadrupole moment of the deflector.
Of course, both
methods give the same results. However, the second method is
more general.

\subsubsection{First Method of Integration}

Let us assume the most general aperiodic form for the time variation 
of the deflector. In linear approximation the
total mass and spin of the deflector are conserved quantities \cite{45} 
so that they do not depend on time at all, 
and we can consider them as
contributing only to the 
static part of the gravitational field of the deflector \cite{45a}. 
The quadrupole moment is not
static. It may be represented through a Fourier transform as
\begin{equation}
{\cal{I}}_{ij}(t-r)=(2\pi)^{-1/2}\displaystyle{\int_{-\infty}^{+\infty}}
\tilde{\cal{I}}_{ij}(\omega)e^{i\omega (t-r)}d\omega,
\label{fur}
\end{equation}
where $\tilde{\cal{I}}_{ij}(\omega)$ is the (complex) 
Fourier image of the quadrupole moment of the deflector
which must be specified for any particular source of 
gravitational waves. Here, we need not know the specific structure of 
$\tilde{\cal{I}}_{ij}(\omega)$ as it will be shown later
it is irrelevant for subsequent calculations. 

Taking time derivatives of the quadrupole moment yields
\begin{equation}
\dot{\cal I}^{ij}=(2\pi)^{-1/2}\displaystyle{\int_{-\infty}^{+\infty}}
(i \omega)\tilde{\cal{I}}_{ij}(\omega)e^{i\omega (t-r)}d\omega,
\label{firs}
\end{equation}
\begin{equation}
\ddot{\cal I}^{ij}=(2\pi)^{-1/2}\displaystyle{\int_{-\infty}^{+\infty}}
(-\omega^2)\tilde{\cal{I}}_{ij}(\omega)e^{i\omega (t-r)}d\omega.
\label{seco}
\end{equation}
Generally speaking, arbitrary aperiodic source of gravitational waves have an
infinite spectrum. However, it is possible to choose that frequency band 
which gives the largest contribution to the spectrum. The mean frequency
$\Omega$ of
this band defines the size of far (wave) zone of the source, as 
being roughly equal
to the wavelength of emitted gravitational waves $\lambda=2\pi c/\Omega$.
For example, if the
deflector of light rays is a binary system, then the strongest
emission of
gravitational waves takes place at twice the mean orbital frequency of the
system. For making estimates we can use the following approximations for
components of the quadrupole moment
\vspace{0.3 cm}
\begin{equation}
\label{etc}
|\dot{\cal I}^{ij}|\simeq \left({\cal M}a\;e\;c\right)\frac{a}{\lambda}\;
,\hspace{1.5 cm}
|\ddot{\cal I}^{ij}|\simeq \left({\cal M}\;e\;c^2\right)
\frac{a^2}{\lambda^2}\;,\hspace{1.5 cm}
etc.\;, 
\end{equation}
where $a$ is a characteristic size of the source of gravitational waves and $e$
is its oblateness, quantifying the deviation of the 
density distribution
from spherical symmetry.

When integrating the equations of light propagation using 
the metric with Fourier 
transform (\ref{fur}) for the quadrupole moment one meets the following
integrals:
\begin{equation}
\label{23}
I_1(\tau,{\bm{\xi}},\omega)={\int_{-\infty}^{\tau}}\frac{\cos[
\omega(\tau-\sqrt{d^2+\tau^2})]}{\sqrt{d^2+\tau^2}}d\tau,
\end{equation}
\begin{equation}
\label{24}
I_2(\tau,{\bm{\xi}},\omega)={\int_{-\infty}^{\tau}}\frac{\sin[
\omega(\tau-\sqrt{d^2+\tau^2})]}
{\sqrt{d^2+\tau^2}}d\tau.
\end{equation} 

In order to evaluate the integrals (\ref{23})-(\ref{24}) it is useful to change
the time argument, $\tau$, to the argument $y$, by the transformation
\begin{equation}
\label{25}
y=\tau-\sqrt{d^2+\tau^2},
\end{equation}
which yields
\begin{equation}
\label{26}
\tau=\frac{y^2-d^2}{2y}, \hspace{1 cm}
\sqrt{d^2+\tau^2}=-\frac{1}{2}\frac{d^2+y^2}{y},
\hspace{1 cm}d\tau=\frac{1}{2}\frac{d^2+y^2}{y^2}dy.
\end{equation}
While the parameter $\tau$ runs from $-\infty$ to $+\infty$, the new 
parameter $y$ runs from $-\infty$ to 0; that is, $y$ is always negative.

After transforming time
arguments, the integrals $I_1$ and $I_2$ are reduced to the cosine- and 
sine integrals respectively (\cite{46}, formula {\bf 8.230}): 
\begin{equation}
\label{27}
I_1(\tau,{\bm{\xi}},\omega)=-{\bf Ci}(\omega y),
\end{equation}
\begin{equation}
\label{28}
I_2(\tau,{\bm{\xi}},\omega)=-{\bf Si}(\omega y),
\end{equation} 
where constants of integration have been omitted. 
Secondary integration of integrals (\ref{27})-(\ref{28}) along the light
trajectory is required as well. Using transformations (\ref{25})-(\ref{26}) 
we obtain
\begin{equation}
\label{30}
J_1(\tau,{\bm{\xi}},\omega)\equiv{\int_{-\infty}^{\tau}}I_1(\tau, {\bm{\xi}}, 
\omega)d\tau=
-\tau\; {\bf Ci}(\omega y)+\frac{1}{2}\omega\;d^2\left[{\bf Si}(\omega y)+
\frac{\cos(\omega y)}{2y}\right]+
\frac{\sin(\omega y)}{2\omega},
\end{equation}
\begin{equation}
\label{29}
J_2(\tau,{\bm{\xi}},\omega)\equiv{\int_{-\infty}^{\tau}}I_2(\tau, {\bm{\xi}}, 
\omega)d\tau=
-\tau\; {\bf Si}(\omega y)+\frac{1}{2}\omega\; d^2\left[{\bf Ci}(\omega y)-
\frac{\sin(\omega y)}{2y}\right]+
\frac{\cos(\omega y)}{2\omega},
\end{equation}
where constants of integration have again been omitted. 

Using the Fourier transform of the quadrupole moment (\ref{fur}) and formulae
(\ref{23}), (\ref{24}), (\ref{27}), (\ref{28}) one calculates the important
integrals
\vspace{0.3 cm}
\begin{eqnarray}
\label{bb}
B_{ij}(\tau,{\bm{\xi}})&\equiv &\displaystyle{\int_{-\infty}^{\tau}}\frac{{\cal I}_{ij}(t-r)}{r}dt
=(2\pi)^{-1/2}
\displaystyle{\int_{-\infty}^{+\infty}}
\tilde{\cal{I}}_{ij}(\omega)e^{i\omega t^{\ast}}\left[
I_1(\tau,{\bm{\xi}},\omega)+i I_2(\tau,{\bm{\xi}},\omega)\right]d\omega\;,
\end{eqnarray}
\vspace{0.3 cm}
\begin{eqnarray}
\label{cc}
C_{ij}(\tau,{\bm{\xi}})&\equiv&\displaystyle{\int_{-\infty}^{\tau}}\frac{\dot{\cal I}_{ij}(t-r)}{r}dt=(2\pi)^{-1/2}
\displaystyle{\int_{-\infty}^{+\infty}}\omega
\tilde{\cal {I}}_{ij}(\omega)e^{i\omega t^{\ast}}\left[
-I_2(\tau,{\bm{\xi}},\omega)+i I_1(\tau,{\bm{\xi}},\omega)\right]d\omega\;,
\end{eqnarray} 
\vspace{0.3 cm}
\begin{eqnarray}
\label{dd}
D_{ij}(\tau,{\bm{\xi}})&\equiv&\displaystyle{\int_{-\infty}^{\tau}}B_{ij}(\tau,{\bm{\xi}})dt=
(2\pi)^{-1/2}
\displaystyle{\int_{-\infty}^{+\infty}}
\tilde{\cal{I}}_{ij}(\omega)e^{i\omega t^{\ast}}\left[
J_1(\tau,{\bm{\xi}},\omega)+i J_2(\tau,{\bm{\xi}},\omega)\right]d\omega\;,
\end{eqnarray}
\vspace{0.3 cm}
\begin{eqnarray}
\label{ee}
E_{ij}(\tau,{\bm{\xi}})&\equiv&\displaystyle{\int_{-\infty}^{\tau}}C_{ij}(\tau,{\bm{\xi}})dt
=(2\pi)^{-1/2}
\displaystyle{\int_{-\infty}^{+\infty}}\omega
\tilde{\cal{I}}_{ij}(\omega)e^{i\omega t^{\ast}}\left[
J_2(\tau,{\bm{\xi}},\omega)-i
J_1(\tau,{\bm{\xi}},\omega)\right]d\omega\;,\\\nonumber
\end{eqnarray}
where $t^{\ast}$ is the moment of closest
approach of the photon to the origin of coordinate system. 
In what follows, we need only 
partial derivatives with respect to the impact parameter
of the integrals (\ref{bb}) - (\ref{ee}). These can be
calculated rather easily. We have, for example,
\vspace{0.3 cm}
\begin{equation}
\label{iju}
{\hat{\partial}}_i I_1(\tau,{\bm{\xi}},\omega)=
\left(y r\right)^{-1}\cos\left(\omega
y\right) \xi^i,\hspace{1.5 cm}{\hat{\partial}}_i I_2(\tau,{\bm{\xi}},\omega)=
\left(y r\right)^{-1}\sin\left(\omega y\right) \xi^i\;,\\\nonumber
\end{equation}
and so on. Thus, making use of the inverse Fourier transform we obtain
\vspace{0.3 cm}
\begin{eqnarray}
\label{cvt}
{\hat{\partial}}_k B_{ij}(\tau,{\bm{\xi}})&=&
\left(y r\right)^{-1}{\cal I}_{ij}(t-r)
 \xi^k,
\end{eqnarray}
\begin{eqnarray}
\label{cds}
{\hat{\partial}}_{\tau} B_{ij}(\tau,{\bm{\xi}})&=&\left(1-\frac{\tau}{r}\right)
\frac{{\cal I}_{ij}(t-r)}{y},
\end{eqnarray}
\begin{eqnarray}
\label{ctui}
{\hat{\partial}}_k C_{ij}(\tau,{\bm{\xi}})&=&\left(y r\right)^{-1}
\dot{\cal I}_{ij}(t-r) \xi^k.
\end{eqnarray}
\begin{eqnarray}
\label{cdo}
{\hat{\partial}}_{\tau} C_{ij}(\tau,{\bm{\xi}})&=&\left(1-\frac{\tau}{r}\right)
\frac{\dot{\cal I}_{ij}(t-r)}{y},
\end{eqnarray}

Calculation of partial derivatives from integrals $D_{ij}(\tau,{\bm{\xi}})$ and 
$E_{ij}(\tau,{\bm{\xi}})$ may be done without  difficulty 
in a similar fashion using equations (\ref{30})-(\ref{29}).

\subsubsection{Second Method of Integration}

The second method also uses the substitutions (\ref{25}), (\ref{26}).
The integrals (\ref{bb}) - (\ref{cc}) are brought into the form
\vspace{0.3 cm}
\begin{eqnarray}
\label{bbq}
B_{ij}(\tau,{\bm{\xi}})&\equiv& \displaystyle{\int_{-\infty}^{\tau}}
\frac{{\cal I}_{ij}(t-r)}{r}dt
=-\displaystyle{\int_{-\infty}^{y}}\frac{{\cal
I}_{ij}(t^{\ast}+\zeta)}{\zeta}d\zeta\;,
\end{eqnarray}
\vspace{0.3 cm}
\begin{eqnarray}
\label{ccq}
C_{ij}(\tau,{\bm{\xi}})&\equiv&\displaystyle{\int_{-\infty}^{\tau}}
\frac{\dot{\cal I}_{ij}(t-r)}{r}dt=
-\displaystyle{\int_{-\infty}^{y}}\frac{\dot{\cal
I}_{ij}(t^{\ast}+\zeta)}{\zeta}d\zeta\;.\\\nonumber
\end{eqnarray}
One sees that the integrands of the integrals do not depend on the 
parameters  $d$ and $\tau$ at all. They are present only in the upper
limit of integration. Hence, the integrals (\ref{bbq}), (\ref{ccq}) are functions
of the variable $y$ only, that is $B_{ij}(\tau,{\bm{\xi}})=B_{ij}(y)$ and 
$C_{ij}(\tau,{\bm{\xi}})=C_{ij}(y)$. Making use of the transformations 
(\ref{25}), (\ref{26}), 
the integrals (\ref{dd}), (\ref{ee}) are reduced to the
expressions
\vspace{0.3 cm}
\begin{equation}
\label{ddq}
D_{ij}(\tau,{\bm{\xi}})\equiv
\displaystyle{\int_{-\infty}^{\tau}}B_{ij}(\tau,{\bm{\xi}})dt=
\frac{1}{2}\displaystyle{\int_{-\infty}^{y}}B_{ij}(\zeta)d\zeta+
\frac{d^2}{2}\displaystyle{\int_{-\infty}^{y}}\frac{B_{ij}(\zeta)}
{\zeta^2}d\zeta,
\end{equation}
\vspace{0.3 cm}
\begin{equation}
\label{eeq}
E_{ij}(\tau,{\bm{\xi}})
\equiv\displaystyle{\int_{-\infty}^{\tau}}C_{ij}(\tau,{\bm{\xi}})dt
=\frac{1}{2}\displaystyle{\int_{-\infty}^{y}}C_{ij}(\zeta)d\zeta+
\frac{d^2}{2}\displaystyle{\int_{-\infty}^{y}}\frac{C_{ij}(\zeta)}
{\zeta^2}d\zeta\;.\\\nonumber
\end{equation}
Hence, the integrals $D_{ij}(\tau,{\bm{\xi}})$, $E_{ij}(\tau,{\bm{\xi}})$ are
also functions of the variable $y$ only. 

We  stress once again that 
our formalism holds true for arbitrary dependence of the quadrupole moment
of the localized source on time, 
and includes the case of sources which produce
bursts of gravitational radiation, such as 
supernova explosions or coalescence of
binary systems, as well as periodic systems.
Indeed, suppose that the burst starts at the
moment $t_1$ and terminates at the moment $t_2$. We assume for simplicity
that before and after the burst the quadrupole moment of the source 
is identically zero.
During the burst, 
the tensor function ${\cal F}_{ij}(t)$
describes the time dependence of the quadrupole moment.
Then all formulae derived in this paper hold, if we
describe the quadrupole moment of the source as a product of two Heaviside step
functions with the tensor function ${\cal F}_{ij}(t)$.
Thus, for any
moment of time we write
\begin{eqnarray}
\label{hev}
{\cal I}_{ij}(t)&=&H(t-t_1)H(t_2-t){\cal F}_{ij}(t)\;,
\end{eqnarray}
where the Heaviside step function is defined as follows
\begin{equation}
\label{det}
H(t-T)=
  \cases{
     1\quad\quad\text{if $t>T$,}\cr
     0\quad\quad\text{otherwise.}\cr
        }
\end{equation}
Time derivatives of the quadrupole moment are calculated 
taking into account that
$\dot{H}(t-T)=\delta(t-T)$ 
is the Dirac delta-function, 
and $\delta(t-t_1){\cal F}_{ij}(t_1)=
\delta(t-t_2){\cal F}_{ij}(t_2)=0$. This yields 
\begin{equation}
\label{differ}
\dot{\cal I}_{ij}(t)=H(t-t_1)H(t_2-t)\dot{\cal F}_{ij}(t)\;,\quad\quad\quad
\ddot{\cal I}_{ij}(t)=H(t-t_1)H(t_2-t)\ddot{\cal F}_{ij}(t)\;,
\end{equation}
and similar formulae for higher derivatives.

It is evident from the structure of integrals (\ref{bbq})-(\ref{eeq}) 
that taking partial derivatives of any of the foregoing 
integrals is reduced to taking the partial derivative with respect to 
$y$. In particular, we obtain
\vspace{0.3 cm}
\begin{eqnarray}
\label{pzk}
{\hat{\partial}}_j B_{pq}(\tau,{\bm{\xi}})&=&-\frac{{\cal
I}_{pq}(t^{\ast}+y)}{y}{\hat{\partial}}_j y=
\left(y r\right)^{-1}{\cal I}_{pq}(t-r) \xi^j\;,\\\nonumber
\end{eqnarray}
which exactly coincides with the result (\ref{cvt}) derived above
using the inverse Fourier transform method.
Second and third partial derivatives of the function $B_{ij}(\tau,{\bm{\xi}})$
with respect to the impact parameter will be useful subsequently. They are
calculated making use of formula (\ref{pzk}). This yields
\vspace{0.3 cm}
\begin{eqnarray}
\label{pzka}
{\hat{\partial}}_{jk} B_{pq}(\tau,{\bm{\xi}})&=&\left(y
r\right)^{-1}\left[P_{jk}+\frac{\xi_j \xi_k}{y r}-
\frac{\xi_j \xi_k}{r^2}\right]{\cal I}_{pq}(t-r)-
\frac{\xi_j \xi_k}{y r^2}\dot{\cal I}_{pq}(t-r)\;,
\end{eqnarray}
\vspace{0.3 cm}
and
\vspace{0.3 cm}
\begin{eqnarray}
\label{pzkab}
{\hat{\partial}}_{ijk} B_{pq}(\tau,{\bm{\xi}})&=&\left(y
r\right)^{-1}\left[\frac{\xi_i P_{jk}}{y r}+\frac{2\xi_k P_{ij}}{y r}+
\frac{2\xi_i \xi_j \xi_k}{y^2 r^2}\right.\nonumber\\ \\\nonumber&&\mbox{}-
\left.\frac{\xi_i P_{jk}}{r^2}-
\frac{2\xi_k P_{ij}}{r^2}-\frac{3\xi_i \xi_j \xi_k}{y r^3}+
\frac{3\xi_i \xi_j \xi_k}{r^4}\right]{\cal I}_{pq}(t-r)\nonumber\\\nonumber
\\\nonumber&&\mbox{}-\left(y
r\right)^{-1}\left[\frac{\xi_i P_{jk}}{r}+\frac{2\xi_k P_{ij}}{ r}+
\frac{2\xi_i \xi_j \xi_k}{y r^2}-
\frac{3\xi_i \xi_j \xi_k}{r^3}\right]\dot{\cal I}_{pq}(t-r)
\nonumber\\\nonumber
\\\nonumber&&\mbox{}+\frac{\xi_i \xi_j \xi_k}{y r^3}\ddot{\cal
I}_{pq}(t-r)\;.\\\nonumber
\end{eqnarray}
We note that the formulae of partial differentiation of
$C_{ij}(\tau,{\bm{\xi}})$ look the same as for $B_{ij}(\tau,{\bm{\xi}})$
after taking into account the fact that the 
integral (\ref{ccq}) depends on the
first time derivative of the quadrupole moment. The derivatives of the 
functionals $E_{ij}(\tau,{\bm{\xi}})$ and  $D_{ij}(\tau,{\bm{\xi}})$ can be
obtained using relationships (\ref{ddq})-(\ref{eeq}) and derviatives of
$B_{ij}(\tau,{\bm{\xi}})$ and $C_{ij}(\tau,{\bm{\xi}})$. For example,
\begin{eqnarray}
\label{diffd1}
{\hat{\partial}}_{j} D_{pq}(\tau,{\bm{\xi}})&=&\xi^j\left[
\frac{B_{pq}(\tau,{\bm{\xi}})}{y}
+\displaystyle{\int_{-\infty}^{y}}\frac{B_{ij}(\zeta)}
{\zeta^2}d\zeta\right]\;,\\\nonumber\\\label{diffd2}
{\hat{\partial}}_{jk}
D_{pq}(\tau,{\bm{\xi}})&=&
\frac{\xi^j}{y}{\hat{\partial}}_k
B_{pq}(\tau,{\bm{\xi}})+P^{jk}\left[
\frac{B_{pq}(\tau,{\bm{\xi}})}{y}
+\displaystyle{\int_{-\infty}^{y}}\frac{B_{ij}(\zeta)}
{\zeta^2}d\zeta\right]\;,\\\nonumber\\\label{diffd3}
{\hat{\partial}}_{ijk}
D_{pq}(\tau,{\bm{\xi}})&=&\frac{1}{y}\left[\left(P^{ij}+\frac{\xi^i\xi^j}{y
r}\right){\hat{\partial}}_k
B_{pq}(\tau,{\bm{\xi}})+P^{jk}{\hat{\partial}}_i
B_{pq}(\tau,{\bm{\xi}})+\xi^j {\hat{\partial}}_{ik}
B_{pq}(\tau,{\bm{\xi}})\right]\;.\\\nonumber
\end{eqnarray} 
It is worth emphasizing that the third partial derivative of 
$D_{pq}(\tau,{\bm{\xi}})$ 
does not include the integral  
$B_{pq}(\tau,{\bm{\xi}})$ by itself, as
might be expected, but only its first and second derivatives. 
Therefore, the third partial derivative of 
$D_{pq}(\tau,{\bm{\xi}})$ does not depend on the past history of propagation of
the light ray (see
formulae (\ref{pzk}) and (\ref{pzka})).

Now, after making these remarks, we are ready to discuss the 
relativistic perturbations of the photon's
trajectory in the radiative gravitational field of a localized source
deflecting light rays.

\section{Perturbations of Photon's Trajectory}

We first note that in terms of the new variables $\tau$ and $\xi^i$ the
components of the ``canonical" 
metric tensor (\ref{7})-(\ref{9}) taken at an 
arbitrary point on the light ray 
can be re-written as follows \cite{47}:
\begin{eqnarray}\label{4aa}
h_{00}^{can.}(\tau,{\bm{\xi}})&=
&\frac{2{\cal M}}{r}+\left({\hat{\partial}}_{ij}+2k_i\hat{\partial}_{j\tau}+k_i
k_j{\hat{\partial}}_{\tau\tau}\right)\left[\frac{{\cal
I}_{ij}(t-r)}{r}\right]-\\\nonumber  \\ \nonumber
& &\mbox{}2\left(k_i {\hat{\partial}}_{j}+k_i k_j{\hat{\partial}}_{\tau}\right)
\left[\frac{ \dot{\cal{I}}_{ij}(t-r)}{r}\right]+k_i k_j
\frac{\ddot{\cal {I}}_{ij}(t-r)}{r}\;,\\\nonumber  \\\label{5aa} 
h_{0i}^{can.}(\tau,{\bm{\xi}})&=&-\frac{2\epsilon_{ipq}{\cal S}^p x^q_N}{r^3}+
2\left({\hat{\partial}}_j+k_j {\hat{\partial}}_{\tau}\right)
\left[\frac{\dot{\cal {I}}_{ij}(t-r)}{r}\right]-
2k_j \frac{\ddot{\cal {I}}_{ij}(t-r)}{r}\;,\\\nonumber  \\\label{6aa}
h_{ij}^{can.}(\tau,{\bm{\xi}})&=&\delta_{ij}
h_{00}^{can.}(\tau,{\bm{\xi}})+\frac{2}{r}\ddot{\cal {I}}_{ij}(t-r),
\end{eqnarray}
where in the right-hand side of all formulae it is implicitly assumed that
variables $t$, $x^i$ are replaced by $\tau$ and $\xi^i$, and 
$\hat{\partial}_i \equiv P_i^j\partial/\partial\xi^j$, 
$\hat{\partial}_{\tau} \equiv \partial/\partial\tau$. In addition, 
 note
that the dot over the quadrupole moment ${\cal I}_{ij}$ 
takes the usual meaning of
differentiation with respect to time, which must be completed first, 
before  
substitution of $t$ and $x^i$ for $\tau$ and $\xi^i$, and before 
taking any other derivative.  

The metric tensor (\ref{4aa}) - (\ref{6aa}) is used in the equations of motion of
light rays (\ref{eqnm}) 
which are reduced with the help of formula (\ref{20}) to the
expression:
\vspace{0.3 cm}
\begin{eqnarray}
\label{zoya}
\ddot{x}^i(\tau)&=&\left[2{\cal M}\left({\hat{\partial}}_{i}-k_i\hat{\partial}_{\tau}\right)-
2{\cal S}^p\left(\epsilon_{ipq}{\hat{\partial}}_{q\tau}+
k_q\epsilon_{ipq}{\hat{\partial}}_{\tau\tau}-k_j\epsilon_{jpq}{\hat{\partial}}_{iq}
\right)\right]\biggl\{\frac{1}{r}\biggr\}+\\\nonumber  \\ \nonumber
& &\mbox{}\left({\hat{\partial}}_{ipq}-k_i{\hat{\partial}}_{pq\tau}+
2k_p{\hat{\partial}}_{iq\tau}+k_p k_q{\hat{\partial}}_{i\tau\tau}-
2k_i k_p{\hat{\partial}}_{q\tau\tau}-k_i k_p
k_q{\hat{\partial}}_{\tau\tau\tau} \right)
\biggl\{\frac{{\cal {I}}_{pq}(t-r)}{r}\biggl\}+\nonumber \\ \nonumber \\
\nonumber & &\mbox{}
2\left(k_i k_p{\hat{\partial}}_{q\tau}-\delta_{ip}{\hat{\partial}}_{q\tau}-
\delta_{ip}k_q{\hat{\partial}}_{\tau\tau}+
k_i k_p k_q{\hat{\partial}}_{\tau\tau}
 \right)\biggl\{\frac{\dot{\cal {I}}_{pq}(t-r)}{r}\biggl\}-
{\hat{\partial}}_{\tau\tau}\left(w^i\;-k^i\;w^0\right)\; ,
\end{eqnarray}
where $w^i$ and $w^0$ are functions given by relationships
(\ref{poh})-(\ref{boh}). Remarkably, no terms depending on the
second time derivatives of the quadrupole moment  appear 
in the
equations of
motion of light rays (\ref{zoya}), because of mutual cancellation. This fact
explicitly demonstrates that gravitational waves emitted by localized sources
are much more elusive from detection by angular deflection
than other authors
suggest. It is worth 
noting that the disappearance of terms with second derivatives
from the quadrupole moment is a local phenomena and is not a result of
integration of equation (\ref{zoya}). This is a characteristic feature of
General Relativity. Alternative theories of gravity do not
possess such a local cancellation of gravitational wave terms. 
This cancellation may be used
for conducting new  tests of General Relativity in the 
weak, radiative
gravitational-field limit. 

Let us
simplify the equations of motion (\ref{zoya}) in order to avoid writing down
cumbersome expressions. We introduce two functions 
$\varphi^i$ and $\varphi^0$ which generate in (\ref{zoya}) 
the time derivatives of second and higher orders. These functions are 
defined:
\vspace{0.3 cm}
\begin{eqnarray}\label{w0}
\varphi^0 &=&-\;2k_p \nabla_q\biggl\{
\frac{{\cal {I}}_{pq}(t-r)}{r}\biggl\}+
k_p k_q\biggl\{\frac{\dot{\cal {I}}_{pq}(t-r)}{r}\biggl\} \;,\\\nonumber\\\label{wi}
\varphi^i &=& 2{\cal S}^p\;k_q\;\epsilon_{ipq}\biggl\{\frac{1}{r}\biggr\}-
k_p\;k_q\nabla_i\biggl\{\frac{{\cal {I}}_{pq}(t-r)}{r}\biggl\}+
2 k_q \biggl\{\frac{\dot{\cal {I}}_{iq}(t-r)}{r}\biggl\}\;,
\end{eqnarray}
where the differential operator $\nabla_i\equiv\partial/\partial x^i$ 
must be applied
before the substitution of the unperturbed trajectory of light rays. 
It can be easily confirmed by straightforward use of formula (\ref{20}) that the
expressions (\ref{w0})-(\ref{wi}) generate terms with second and 
third derivatives with respect to $\tau$ in (\ref{zoya}). The equations 
for the path of the light
ray now assume the form:
\begin{eqnarray}\label{zoya1}
\ddot{x}^i(\tau)&=&\left[2{\cal M}\left({\hat{\partial}}_{i}-
k_i\hat{\partial}_{\tau}\right)-
2{\cal S}^p\left(\epsilon_{ipq}{\hat{\partial}}_{q\tau}
-k_j\epsilon_{jpq}{\hat{\partial}}_{iq}
\right)\right]\biggl\{\frac{1}{r}\biggr\}+\\\nonumber  \\ \nonumber
& &\mbox{}\left({\hat{\partial}}_{ipq}-k_i{\hat{\partial}}_{pq\tau}+
2k_p{\hat{\partial}}_{iq\tau}\right)
\biggl\{\frac{{\cal {I}}_{pq}(t-r)}{r}\biggl\}-
2P_{ij}{\hat{\partial}}_{q\tau}\biggl\{\frac{\dot{\cal {I}}_{jq}(t-r)}{r}\biggl\}-
\\\nonumber  \\ \nonumber\mbox{}&&
{\hat{\partial}}_{\tau\tau}\left[w^i+\varphi^i
-k^i\;\left(w^0+\varphi^0\right)\right]\;.\\\nonumber
\end{eqnarray}
We note that the terms
$\varphi^0$ and $\varphi^i$ are gauge-dependent and can be, in principle,
eliminated from the equations of motion (\ref{zoya}) by choosing appropriate 
gauge functions $w^0$ and $w^i$. However, such a procedure
will introduce a reference system with a coordinate grid  very 
sensitive to the direction to a specific source of light rays; 
that is, to the
vector $k^i$. The coordinate system obtained in this way will be of trifling
practical usage. For this reason we do not recommend the elimination of
functions $\varphi^0$ and $\varphi^i$ from (\ref{zoya}) and give preference
to the ADM-harmonic coordinate system, which admits a much simpler and 
unique 
treatment of observable effects. Thus, we leave the functions 
$\varphi^0$ and $\varphi^i$ in the equations of motion of light rays, 
where gauge
functions $w^0$ and $w^i$ are defined by formulae (\ref{ttt})-(\ref{kkk}).

Proceeding further in this way
and integrating equations (\ref{zoya}) one obtains
\vspace{0.3 cm}
\begin{eqnarray}
\label{jja}
\dot{x}^i(\tau)&=&k^i+\dot{\Xi}^i(\tau)\\\nonumber\\
\label{epr}
x^i(\tau)&=&x^i_N(\tau)+\Xi^i(\tau)-\Xi^i(\tau_0)\;,
\end{eqnarray}
where the unperturbed trajectory of light ray $x^i_N(\tau)$ is determined by
the expression (\ref{17a}). The 
relativistic perturbations to the trajectory are:
\vspace{0.5 cm}
\begin{eqnarray}
\label{aop}
\dot{\Xi}^i(\tau) &=&
\left(2{\cal M}{\hat{\partial}}_{i}+2{\cal
S}^pk_j\epsilon_{jpq}{\hat{\partial}}_{iq}\right)A(\tau,{\bm{\xi}})+
{\hat{\partial}}_{ipq}B_{pq}(\tau,{\bm{\xi}})-
\\ \nonumber \\&&\mbox{}
\left(2{\cal M}k_i + 2{\cal S}^p\epsilon_{ipq}{\hat{\partial}}_q\right)
\biggl\{\frac{1}{r}\biggr\}-
\left(k_i{\hat{\partial}}_{pq}-
2k_p{\hat{\partial}}_{iq}\right)
\biggl\{\frac{{\cal {I}}_{pq}(t-r)}{r}\biggl\}-
2P_{ij}{\hat{\partial}}_{q}\biggl\{\frac{\dot{\cal {I}}_{jq}(t-r)}{r}\biggl\}-
\nonumber\\ \nonumber \\\nonumber&&\mbox{}
{\hat{\partial}}_{\tau}\left[w^i+\varphi^i
-k^i\;\left(w^0+\varphi^0\right)\right]\;,\\ \nonumber \\\nonumber\\
\label{jjj}
\Xi^i(\tau)&=&\left(2{\cal M}{\hat{\partial}}_{i}+2{\cal
S}^pk_j\epsilon_{jpq}{\hat{\partial}}_{iq}\right)B(\tau,{\bm{\xi}})-
\left(2{\cal M}k_i-2{\cal S}^p\epsilon_{ipq}{\hat{\partial}}_q
\right)A(\tau,{\bm{\xi}})
+\\\nonumber 
\\&&\mbox{}{\hat{\partial}}_{ipq}D_{pq}(\tau,{\bm{\xi}})-
\left(k_i{\hat{\partial}}_{pq}-2k_p{\hat{\partial}}_{iq}
\right)B_{pq}(\tau,{\bm{\xi}})
-2P_{ij}{\hat{\partial}}_{q}C_{jq}(\tau,{\bm{\xi}})-
\nonumber\\\nonumber 
\\&&\mbox{}
-w^i(\tau,{\bm{\xi}})-\varphi^i(\tau,{\bm{\xi}})
+k^i\;\left[w^0(\tau,{\bm{\xi}})+\varphi^0(\tau,{\bm{\xi}})\right]\;.
\nonumber\\ \nonumber 
\end{eqnarray}
We emphasize that before differentiation with respect to time $\tau$
or impact parameter $\xi^i$, 
one has to differentiate the quadrupole moment with
respect to time $t$ and make the substitutions: $t\mapsto \tau$,
$r\mapsto \sqrt{d^2+\tau^2}$, $r_0\mapsto \sqrt{d^2+\tau_0^2}$. We also wish
to underline that the only integrals which need be calculated
explicitly in expressions (\ref{aop})-(\ref{jjj}) are $A(\tau,{\bm{\xi}})$ and
$B(\tau,{\bm{\xi}})$. All other
integrals are acted upon by
partial derivatives, which reduce them to ordinary
functions as  explained in the 
previous section. This remarkable fact allows
considerable simplification of 
the calculations. 
This simplification results from the fact that the
integrands can be formed from retarded potentials independent
of impact parameter, after using the
transformation of variables (\ref{25}).
This would be impossible if the metric tensor were
not a function of retarded time
$t-r$. Thus,  retardation simplifies the calculations in the case of
time-dependent gravitational fields.
In the case of a static or stationary 
gravitational field, the calculation of
propagation of light can be done using the same technique since one
can always consider a constant multipole also as a (constant) function of
retarded time. For this reason, more involved calculations of light propagation
(e.g. see \cite{24} and \cite{32}) can be simplified as well.     

The functions $w^i$ and $w^0$, which describe 
freedom in choosing coordinate systems, are
taken from formulae (\ref{ttt})-(\ref{kkk}) of Appendix B. Consequently, the
integrals of equations of light propagation (\ref{zoya}) 
expressed
in the ADM-harmonic coordinate gauge possess a simple interpretation of
observable effects, as discussed in the following section. 

It is convenient to obtain an 
expression for unit vector $k^i$ written in terms
of spatial coordinates of the 
points of emission, ${\bf x_0}$, and observation, 
${\bf x}$, of the light ray. From formula (\ref{epr}) one has
\begin{eqnarray}
\label{uuu}
k^i &=&-
K^i-\frac{P_j^i\left[\Xi^j
(\tau,{\bm{\xi}})-\Xi^j(\tau_0,{\bm{\xi}})\right]}{|{\bf x}-{\bf
x}_0|}\;,\\\nonumber
\end{eqnarray}
or more explicitly
\begin{eqnarray}
\label{expli}
k^i
&=&-K^i-\beta^i(\tau,{\bm{\xi}})+\beta^i(\tau_0,{\bm{\xi}})\;,
\\\nonumber\\
\beta^i(\tau,{\bm{\xi}})&=&\beta^i_M(\tau,{\bm{\xi}})+
\beta^i_S(\tau,{\bm{\xi}})+\beta^i_Q(\tau,{\bm{\xi}})\;,\\\nonumber
\end{eqnarray}
where the relativistic corrections $\beta^i(\tau,{\bm{\xi}})$ 
to the vector $K^i$ are defined as follows:\vspace{0.5
cm}
\begin{eqnarray}
\label{corr}
\beta^i_M(\tau,{\bm{\xi}})&=&\frac{
2{\cal M}{\hat{\partial}}_{i}B(\tau,{\bm{\xi}})}{|{\bf x}-{\bf x}_0|}\;,
\\\nonumber\\
\label{correc}
\beta^i_S(\tau,{\bm{\xi}})&=&\frac{2{\cal
S}^pk_j\epsilon_{jpq}{\hat{\partial}}_{iq}B(\tau,{\bm{\xi}})+
2P^{ij}{\cal S}^p\epsilon_{jpq}{\hat{\partial}}_q A(\tau,{\bm{\xi}})}
{|{\bf x}-{\bf x}_0|}\;,\\\nonumber\\
\label{coper}
\beta^i_Q(\tau,{\bm{\xi}})&=&\frac{
{\hat{\partial}}_{ipq}D_{pq}(\tau,{\bm{\xi}})+2k_p{\hat{\partial}}_{iq}
B_{pq}(\tau,{\bm{\xi}})
-2P^{ij}{\hat{\partial}}_{q}C_{jq}(\tau,{\bm{\xi}})
-P^i_j\left[w^j(\tau,{\bm{\xi}})+\varphi^j(\tau,{\bm{\xi}})\right]}
{|{\bf x}-{\bf x}_0|}
\;.
\\ \nonumber 
\end{eqnarray}
The relativistic corrections $\beta^i(\tau_0,{\bm{\xi}})$ are obtained 
by  replacing the parameter $\tau$ in the numerators
of expressions (\ref{corr})-(\ref{coper}) by $\tau_0$. One notes that in
equation (\ref{expli})
the unit Euclidean vector 
\begin{eqnarray}
\label{unitv}
K^i=-\frac{x^i-x_0^i}{|{\bf x}-{\bf x}_0|}\;
\end{eqnarray}
defines the direction from the observer towards the source of light and may be 
interpreted as a
direction in  asymptotically flat space-time \cite{48}.
Relationship (\ref{uuu}) allows
us
to apply the results of integration of equation of light propagation 
to the boundary value problem as well. 
The boundary value problem is formulated
in terms of initial ${\bf x}_0$ and final ${\bf x}$ positions of the photon
\begin{equation}\label{bvp}
{\bf x}(t)={\bf x}\;,\quad\quad {\bf x}(t_0)={\bf x}_0\;,
\end{equation}
whilst the initial-boundary value problem (\ref{1}) is formulated by means
of assignment of the initial position ${\bf x}_0$ and velocity of photon 
at past null infinity. 
The relativistic correction to the vector
$K^i$ contains in its denominator the 
large numerical value of the distance
between observer and source of light. However, the difference $\Xi^j
(\tau,{\bm{\xi}})-\Xi^j(\tau_0,{\bm{\xi}})$ in the numerator of (\ref{uuu}) may be
of the same order as $|{\bf x}-{\bf x}_0|$ itself. 
For this reason the relativistic correction in question 
must be 
taken into account, in general, for
calculation of  light deflection in the cases of 
finite distances of
observer or source of light from the localized source of gravitational
waves. Only in the case where observer and source of light reside 
at large distances on 
opposite
sides of the source of gravitational waves, 
as  was assumed in the paper by Damour \&
Esposito-Far\'ese (1998), can the relativistic correction $\beta^i$ 
beneglected.  

\section{Basic Observable Relativistic Effects}

\subsection{Time Delay}

The gravitational time delay is derived from equations (\ref{epr}), (\ref{jjj}). In
order to obtain the expression for the time delay we multiply equation 
(\ref{epr}) by itself and then find the 
difference $t-t_0$ by taking the 
square root and
using an expansion with respect to small relativistic parameters. This yields:
\vspace{0.3 cm}
\begin{eqnarray}
t-t_0&=&|{\bf x}-{\bf x}_0|-{\bf{k}}\cdot{\bm{\Xi}}(\tau)+
{\bf{k}}\cdot{\bm{\Xi}}(\tau_0)\;,\\\nonumber
\end{eqnarray}
or\vspace{0.3 cm}
\begin{eqnarray}
\label{qer}
t-t_0&=&|{\bf x}-{\bf x}_0|+\Delta_M(t,t_0)+\Delta_S(t,t_0)+
\Delta_Q(t,t_0),
\end{eqnarray}
where $|{\bf x}-{\bf x}_0|$ is the usual Euclidean distance 
\cite{49} 
between the points
of emission, ${\bf x}_0$, and reception, ${\bf x}$, of the photon, $\Delta_M$ 
is the classical
Shapiro delay produced by the (constant) spherically symmetric part of 
the gravitational
field of the deflector, $\Delta_S$ is the Lense-Thirring or Kerr delay due to
the (constant) spin of the localized source of gravitational waves,
and  $\Delta_Q$ describes an additional
delay caused by the time dependent quadrupole moment of the source. 
Specifically we obtain:
\vspace{0.3 cm}
\begin{eqnarray}
\label{mass1}
\Delta_M=2{\cal M} \ln\left[\frac{r+\tau}{r_0+\tau_0}\right]
\end{eqnarray} 
\vspace{0.3 cm}
\begin{eqnarray}
\label{spin1}
\Delta_S&=&-2\epsilon_{ijk}k^j{\cal S}^k {\hat{\partial}}_i 
\ln\left[\frac{r+\tau}{r_0+\tau_0}\right]
\end{eqnarray}
\vspace{0.3 cm}
\begin{eqnarray}
\label{quad1}
\Delta_Q&=&{\hat{\partial}}_{ij}
\left[B_{ij}(\tau,{\bm{\xi}})-B_{ij}(\tau_0,{\bm{\xi}})\right]+
\delta_Q(\tau,{\bm{\xi}})-\delta_Q(\tau_0,{\bm{\xi}})\;,
\end{eqnarray}
where
\begin{eqnarray}
\label{delta}
\delta_Q(\tau,{\bm{\xi}}) &=& k^i\left(w^i+\varphi^i\right)-
w^0-\varphi^0=\\\nonumber\\\nonumber\mbox{}&&
\frac{1}{2}{\hat{\partial}}_{\tau}\left[\nabla_p\nabla_q\biggl\{
\frac{^{(-2)}{\cal{I}}_{pq}(t-r)}{r}\biggl\}\right]-
\nabla_p\nabla_q\biggl\{
\frac{^{(-1)}{\cal{I}}_{pq}(t-r)}{r}\biggl\}-
\\\nonumber\\\nonumber\\\nonumber\mbox{}&&
k_p k_q{\hat{\partial}}_{\tau}\biggl\{\frac{{\cal{I}}_{pq}(t-r)}{r}\biggl\}+
2k_p k_q\biggl\{\frac{\dot{\cal{I}}_{pq}(t-r)}{r}\biggl\}\;.\\\nonumber
\end{eqnarray}
and functions $^{(-1)}{\cal{I}}_{pq}(t-r)$ and $^{(-2)}{\cal{I}}_{pq}(t-r)$
are defined by formula (\ref{integr}) of Appendix B.
The expression for the second derivative of function $B_{ij}(\tau,{\bm{\xi}})$ has
been given in equation (\ref{pzka}). 
The other derivatives appearing in $\Delta_Q$ are
as follows
\vspace{0.3 cm}
\begin{eqnarray}
\label{dertau}
{\hat{\partial}}_{\tau}\biggl\{\frac{{\cal I}_{ij}(t-r)}{r}\biggr\}&=&
-\frac{y}{r}\frac{\dot{\cal I}_{ij}(t-r)}{r}-
\frac{\tau}{r}\frac{{\cal I}_{ij}(t-r)}{r^2}\;,\\\nonumber\\\nonumber\\
\label{oij}
\nabla_p\nabla_q\biggl\{
\frac{^{(-1)}{\cal{I}}_{pq}(t-r)}{r}\biggl\} &=&
\left[\dot{\cal I}_{pq}(t-r)+
3\frac{{\cal{I}}_{pq}(t-r)}{r}+3\frac{^{(-1)}{\cal{I}}_{pq}(t-r)}{r^2}
\right]\frac{x^p_N x^q_N}{r^3}\;,\\\nonumber\\\nonumber\\
\label{mnb}
\nabla_p\nabla_q\biggl\{
\frac{^{(-2)}{\cal{I}}_{pq}(t-r)}{r}\biggl\} &=&
\left[{\cal I}_{pq}(t-r)+
3\frac{^{(-1)}{\cal{I}}_{pq}(t-r)}{r}+3\frac{^{(-2)}{\cal{I}}_{pq}(t-r)}{r^2}
\right]\frac{x^p_N x^q_N}{r^3}\;,\\\nonumber\\\nonumber\\
\label{dertxx}
{\hat{\partial}}_{\tau}\biggl\{\nabla_p\nabla_q
\frac{^{(-2)}{\cal{I}}_{pq}(t-r)}{r}\biggl\} &=&
2\left[{\cal I}_{pq}(t-r)+
3\frac{^{(-1)}{\cal{I}}_{pq}(t-r)}{r}+3\frac{^{(-2)}{\cal{I}}_{pq}(t-r)}{r^2}
\right]\frac{x^q_N k^p }{r^3}\\\nonumber\\\nonumber
\mbox{}&&
-3\frac{\tau}{r}\left[{\cal I}_{pq}(t-r)+4\frac{
^{(-1)}{\cal{I}}_{pq}(t-r)}{r}+5\frac{^{(-2)}{\cal{I}}_{pq}(t-r)}{r^2}\right]
\frac{x^p_N x^q_N}{r^4}\\\nonumber\\\nonumber
\mbox{}&&
-\frac{y}{r}\left[\dot{\cal I}_{pq}(t-r)+
3\frac{{\cal{I}}_{pq}(t-r)}{r}+3\frac{^{(-1)}{\cal{I}}_{pq}(t-r)}{r^2}
\right]\frac{x^p_N x^q_N}{r^3}\;.
\end{eqnarray}

  
The relationship (\ref{qer}) for the time delay has been 
derived with respect to 
coordinate time $t$. In order to convert this relationship
to observable proper time, we
assume for simplicity that the observer is in a state of free fall
and that his velocity is negligibly small at the point of 
observation, with spatial
coordinate ${\bf x}$. 
If the observer's velocity is not small 
an additional
Lorentz transformation of time must be applied. Transformation from the
ADM-harmonic coordinate time $t$ to proper time $T$ is made with 
the help of the formula
(e.g. see \cite{50})
\vspace{0.3 cm}
\begin{eqnarray}
\label{prop}
dT&=&dt\sqrt{- g_{00}(t,{\bf x})}=dt\left(1-\frac{1}{2}h_{00}\right)\;.
\end{eqnarray} 
Implementation of formula (\ref{adm1}) for $h_{00}$ and subsequent
integration of (\ref{prop}) with respect to time  yields
\vspace{0.3 cm}
\begin{eqnarray}
\label{tra}
T&=&\left(1-\frac{{\cal M}}{r}\right)\left(t-t_i\right)\;,
\end{eqnarray}
where $t_{\rm i}$ is the initial epoch of observation and all
velocity-dependent terms are assumed small, as
argued above, and are therefore omitted.
We also  stress that under usual
circumstances the distance $r$ is so large that the difference between 
the observer's proper 
time
 and coordinate time can be neglected. Thus, we are allowed to treat
coordinate time $t$ as proper time.

We note that the time delay in the propagation of light 
depends not only on instantaneous functions of
retarded time but also on the integrals of time $^{(-1)}{\cal{I}}_{pq}(t-r)$ and 
$^{(-2)}{\cal{I}}_{pq}(t-r)$.
These integrals describe the whole past history of the source
of gravitational waves up to the moment of observation. Under usual
circumstances, the influence of 
such integrals on the time delay is expected to
be small. However, this question deserves more detailed discussion and will be
studied in more detail elsewhere. 
For example, these terms  may be revealed
in observations as the ``kinematic resonance effect'' predicted by 
Braginsky \& Grishchuk \cite{14}. These terms may be also important for detection
of the solar $g$-mode tidal oscillations by the LISA gravitational- wave antenna in
space \cite{51}.    

\subsection{Deflection of Light}

The coordinate direction to the source of light measured at the point 
of observation ${\bf x}$ is
defined by the four-vector $p^\alpha=(1,p^i)$ where $p^i=-\dot{x}^i$, or
\vspace{0.3 cm}
\begin{eqnarray}
\label{coor}
p^i&=&-k^i-\dot{\Xi}^i\;,
\end{eqnarray} 
and  the minus sign directs the vector $p^i$ 
from  observer to the source of light. However, the coordinate 
direction $p^i$ is not a directly observable quantity. A real observable vector 
towards the source of light, $s^\alpha=(1,s^i)$, is defined with respect to the local
inertial frame of the observer. 
In this frame $s^i=-dX^i/dT$, where $T$ is the
observer's proper time and $X^i$ are spatial coordinates of the local inertial 
frame. We shall assume for simplicity that observer is at 
rest \cite{52} with respect to
the (global) ADM-harmonic coordinate system $(t,x^i)$. Then the infinitesimal
transformation from the global ADM-harmonic coordinates $(t,x^i)$ to the local
coordinates $(T,X^i)$ is given by the formulas 
\vspace{0.3 cm}
\begin{eqnarray}
\label{trans}
dT=\Lambda^0_0\; dt+\Lambda^0_j\; dx^j\;\;\;&,&\;\;
dX^i=\Lambda^i_0\; dt+\Lambda^i_j\; dx^j\;\;\;,
\end{eqnarray}
where the matrix of transformation $\Lambda^{\alpha}_{\beta}$ is defined by the
requirements of orthonormality 
\vspace{0.3 cm}
\begin{eqnarray}
\label{ort}
g_{\alpha\beta}&=&\eta_{\mu\nu}\Lambda^{\mu}_{\alpha}\Lambda^{\nu}_{\beta}\;.
\end{eqnarray}
In particular, the orthonormality condition (\ref{ort}) assumes that spatial 
angles and lengths at the point of observations are measured with the 
Euclidean metric $\delta_{ij}$. Because the vector $s^\alpha$ is 
isotropic, we conclude that the Euclidean length $|{\bf s}|$ of 
the vector $s^i$ 
is equal to 1. Indeed, one has 
\begin{eqnarray}
\label{unity}
\eta_{\alpha\beta}s^\alpha s^\beta&=&-1+{\bf s}^2=0\;.
\end{eqnarray}
Hence, $|{\bf s}|=1$.

In the linear approximation with respect to G, 
the matrix of the transformation is 
as follows \cite{31}  
\vspace{0.3 cm}
\begin{eqnarray}
\label{lambda}
\Lambda^0_0&=&1-\frac{1}{2}h_{00}(t,{\bf x})\;,\nonumber\\
\mbox{} \Lambda^0_i&=&-h_{0i}(t,{\bf x})\;,\nonumber\\ 
\mbox{} \Lambda^i_0&=&0\;,\nonumber\\\mbox{} 
 \Lambda^i_j&=&\left[1+\frac{1}{2}h_{00}(t,{\bf x})\right]\delta_{ij}+
 \frac{1}{2}h^{TT}_{ij}(t,{\bf x})\;.
\end{eqnarray}
Using the transformation (\ref{trans}) we obtain the relationship between the 
observable vector $s^i$ and the coordinate direction $p^i$
\vspace{0.3 cm}
\begin{eqnarray}
\label{rls}
s^i&=&-\frac{\Lambda^i_0-\Lambda^i_j\; p^j}{\Lambda^0_0-\Lambda^0_j\; p^j}\;.
\end{eqnarray}
In the linear approximation this takes the form
\vspace{0.3 cm}
\begin{eqnarray}
\label{form}
s^i&=&
\left(1+h_{00}-h_{0j}p^j\right)p^i+\frac{1}{2}h^{TT}_{ij}p^j\;.
\end{eqnarray}
Remembering that vector $|{\bf s}|=1$, 
we find the Euclidean norm of the 
vector $p^i$ from the relationship 
\vspace{0.3 cm}
\begin{eqnarray}
\label{norma}
|{\bf p}|&=&1-h_{00}+h_{0j}p^j-\frac{1}{2}h^{TT}_{ij}p^i p^j\;,
\end{eqnarray}
which brings equation (\ref{form}) to the form
\begin{eqnarray}
\label{bnm}
s^i&=&m^i+\frac{1}{2}P^{ij}m^q h^{TT}_{jq}(t,{\bf x})\;,
\end{eqnarray}
where the Euclidean unit vector $m^i=p^i/|{\bf p}|$.

Let us now denote 
by $\alpha^i$ the dimensionless vector describing the total angle of 
deflection of the light ray measured at the point of observation, 
and calculated 
with respect to
vector $k^i$ given at  past null infinity. It is defined according
to the relationship \cite{32}
\vspace{0.3 cm}
\begin{eqnarray}
\alpha^i(\tau,{\bm{\xi}})&=&k^i [{\bf k}\cdot 
\dot{\bm{\Xi}}(\tau,{\bm{\xi}})]-\dot{\Xi}^i(\tau,{\bm{\xi}})\;,
\end{eqnarray}
or
\begin{eqnarray}
\label{ang}
\alpha^i(\tau,{\bm{\xi}})&=&-\;P^i_j\;\dot{\Xi}^j(\tau,{\bm{\xi}})\;.
\end{eqnarray}
As a consequence of the definitions (\ref{coor}) and (\ref{ang}) we conclude 
that
\begin{eqnarray}
\label{uio}
m^i&=&-k^i+\alpha^i(\tau,{\bm{\xi}})\;.
\end{eqnarray}
Accounting for expressions (\ref{bnm}), (\ref{uio}), and (\ref{expli}) we 
obtain for the observed direction to the source of light
\begin{eqnarray}
\label{dop}
s^i(\tau,{\bm{\xi}})&=&K^i+\alpha^i(\tau,{\bm{\xi}})+\beta^i(\tau,{\bm{\xi}})-
\beta^i(\tau_0,{\bm{\xi}})+\gamma^i(\tau,{\bm{\xi}})\;,
\end{eqnarray}
where relativistic corrections $\beta^i$ are defined by equations
(\ref{corr})-(\ref{coper}) and the perturbation
\vspace{0.3 cm} 
\begin{eqnarray}
\label{gamma}
\gamma^i(\tau,{\bm{\xi}})&=&-\frac{1}{2}P^{ij}k^q h^{TT}_{jq}(t,{\bf x})\;.
\end{eqnarray}
If two sources of light (quasars) are observed along the directions $s_1^i$ and 
$s_2^i$ the measured angle $\psi$ between them
in the local inertial frame is:
\vspace{0.3 cm}
\begin{eqnarray}
\label{lkj}
\cos\psi&=&{\bf s}_1\cdot{\bf s}_2\;,
\end{eqnarray}
where the dot denotes the 
usual Euclidean scalar product. It is worth emphasizing
that the observed direction to the source of light includes the relativistic
deflection of the light ray. This depends not only on quantities 
at the point of observation but also on 
$\beta^i(\tau_0,{\bm{\xi}})$, at the point of emission of light. This
remark reveals that according to relation (\ref{lkj}) a single
gravitational wave signal may cause
different  angular displacements for 
different sources of light located at different distances from the source
of gravitational waves. 

Without going into further 
details of the observational procedure we give an explicit
expression for the angle $\alpha^i$. We have
\vspace{0.3 cm}
\begin{eqnarray}
\label{brr}
\alpha^i(\tau,{\bm{\xi}})&=&\alpha_M^i(\tau,{\bm{\xi}})+
\alpha_S^i(\tau,{\bm{\xi}})+\alpha_Q^i(\tau,{\bm{\xi}})\;,\\\nonumber
\end{eqnarray}
where
\vspace{0.3 cm}
\begin{eqnarray}
\label{sdr}
\alpha_M^i(\tau,{\bm{\xi}})&=&
-\;2{\cal M}{\hat{\partial}}_{i}A(\tau,{\bm{\xi}})\;,\\\nonumber\\
\label{dfg}
\alpha_S^i(\tau,{\bm{\xi}})&=&-\;2{\cal
S}^pk_j\epsilon_{jpq}{\hat{\partial}}_{iq}A(\tau,{\bm{\xi}})+
2{\cal S}^p
\left(
P^{ij}\epsilon_{jpq}{\hat{\partial}}_q
+k_q\epsilon_{ipq}{\hat{\partial}}_{\tau}\right)
\biggl\{\frac{1}{r}\biggr\}\;,\\\nonumber\\
\label{dyw}
\alpha_Q^i(\tau,{\bm{\xi}})&=&-\;{\hat{\partial}}_{ipq}B_{pq}(\tau,{\bm{\xi}})-
P_{ij}\left(
2k_p{\hat{\partial}}_{jq}
+k_p k_q{\hat{\partial}}_{j\tau}+2k_p\delta_{jq}{\hat{\partial}}_{\tau\tau}+
2\delta_{jq}{\hat{\partial}}_{p\tau}\right)\times\\
\nonumber\\\nonumber\mbox{}&&
\times\biggl\{\frac{{\cal {I}}_{pq}(t-r)}{r}\biggl\}
+2P_{ij}\left({\hat{\partial}}_{q}
+2k_q{\hat{\partial}}_{\tau}
\right)
\biggl\{\frac{\dot{\cal {I}}_{jq}(t-r)}{r}\biggl\}+
\frac{1}{2}{\hat{\partial}}_{i\tau}\left[\nabla_p\nabla_q\biggl\{
\frac{^{(-2)}{\cal {I}}_{pq}(t-r)}{r}\biggl\}\right]\;.
\end{eqnarray}
The expression for the third spatial derivative of function $B_{pq}(\tau,{\bf
\xi})$ has been given in equation (\ref{pzkab}). The other relevant 
derivatives are:
\vspace{0.3 cm}
\begin{eqnarray}
\label{dotxi}
{\hat{\partial}}_j\biggl\{\frac{\dot{\cal I}_{ij}(t-r)}{r}\biggr\}&=&-\xi^j\left[
\frac{\ddot{\cal I}_{ij}(t-r)}{r^2}+\frac{\dot{\cal I}_{ij}(t-r)}{r^3}\right]\;,
\vspace{0.3 cm}\\\nonumber\\\nonumber\\
\label{dottau}
{\hat{\partial}}_{\tau}\biggl\{\frac{\dot{\cal I}_{ij}(t-r)}{r}\biggr\}&=&
-\frac{y}{r}\frac{\ddot{\cal I}_{ij}(t-r)}{r}-
\frac{\tau}{r}\frac{\dot{\cal I}_{ij}(t-r)}{r^2}\;,
\vspace{0.3 cm}\\\nonumber\\\nonumber\\
\label{dram}
{\hat{\partial}}_{iq}\biggl\{\frac{{\cal
I}_{pq}(t-r)}{r}\biggr\}&=&-P_{iq}\left[\frac{\dot{\cal I}_{pq}(t-r)}{r^2}+
\frac{{\cal I}_{pq}(t-r)}{r^3}\right]+\\\nonumber \\\nonumber&&\mbox{}
\xi_i \xi_q \left[
\frac{\ddot{\cal I}_{pq}(t-r)}{r^3}+\frac{3\dot{\cal I}_{pq}(t-r)}{r^4} +
\frac{3{\cal I}_{pq}(t-r)}{r^5}\right],
\vspace{0.3 cm}\\\nonumber\\\nonumber\\
\label{eris}
{\hat{\partial}}_{i\tau}\biggl\{\frac{{\cal
I}_{pq}(t-r)}{r}\biggr\}&=&
\frac{y}{r}
\left[\frac{\ddot{\cal I}_{pq}(t-r)}{r^2}+
\frac{\dot{\cal I}_{pq}(t-r)}{r^3}\right]\xi_i+
\\\nonumber\\\nonumber\mbox{}&&
\frac{\tau}{r}\left[2\frac{\dot{\cal I}_{pq}(t-r)}{r^3}+
3\frac{{\cal I}_{pq}(t-r)}{r^4}\right]\xi_i\;,\\\nonumber\\\nonumber\\
\label{lls}
{\hat{\partial}}_{\tau\tau}\biggl\{\frac{{\cal
I}_{pq}(t-r)}{r}\biggr\}&=&
\frac{y^2}{r^2}\frac{\ddot{\cal I}_{pq}(t-r)}{r}+
\left(\frac{2y\tau}{r^2}-1\right)\frac{\dot{\cal I}_{pq}(t-r)}{r^2}+
\left(\frac{3\tau^2}{r^2}-1\right)\frac{{\cal I}_{pq}(t-r)}{r^3}\;.
\\\nonumber
\end{eqnarray}
Straightforward but tedious calculation
of the last term in equation (\ref{dyw})
yields\vspace{0.5 cm}
\begin{eqnarray}
\label{tblo}
\hspace{-1 cm}{\hat{\partial}}_{i}\biggl\{\nabla_p\nabla_q
\frac{^{(-2)}{\cal I}_{pq}(t-r)}{r}\biggr\}&=&
2\left[{\cal I}_{pq}(t-r)+
3\frac{^{(-1)}{\cal{I}}_{pq}(t-r)}{r}+3\frac{^{(-2)}{\cal{I}}_{pq}(t-r)}{r^2}
\right]\frac{x^p_N P_{iq} }{r^3}
\\\nonumber\\\nonumber\mbox{}&&\hspace{-2 cm}
-\left[\dot{\cal I}_{pq}(t-r)+
6\frac{{\cal{I}}_{pq}(t-r)}{r}+
15\frac{^{(-1)}{\cal{I}}_{pq}(t-r)}{r^2}+
15\frac{^{(-2)}{\cal{I}}_{pq}(t-r)}{r^3}\right]
\frac{x^p_N x^q_N \xi^i}{r^4}\;.\\\nonumber
\end{eqnarray}
and\vspace{0.5 cm}
\begin{eqnarray}
\label{ted}
\hspace{-1 cm}{\hat{\partial}}_{i\tau}\biggl\{\nabla_p\nabla_q
\frac{^{(-2)}{\cal I}_{pq}(t-r)}{r}\biggr\}&=&
2\left[{\cal I}_{pq}(t-r)+
3\frac{^{(-1)}{\cal{I}}_{pq}(t-r)}{r}+3\frac{^{(-2)}{\cal{I}}_{pq}(t-r)}{r^2}
\right]\frac{k^p P_{iq} }{r^3}\\\nonumber\\\nonumber\mbox{}&&\hspace{-0.5 cm}
-6\frac{\tau}{r}\left[{\cal I}_{pq}(t-r)+4\frac{
^{(-1)}{\cal{I}}_{pq}(t-r)}{r}+5\frac{^{(-2)}{\cal{I}}_{pq}(t-r)}{r^2}\right]
\frac{P_{iq} x^p_N}{r^4}\\\nonumber\\\nonumber\mbox{}&&
-2\frac{y}{r}\left[\dot{\cal I}_{pq}(t-r)+
3\frac{{\cal{I}}_{pq}(t-r)}{r}+3\frac{^{(-1)}{\cal{I}}_{pq}(t-r)}{r^2}
\right]\frac{k^p P_{iq} }{r^3}\\\nonumber\\\nonumber\mbox{}&&\hspace{-2 cm}
-2\left[\dot{\cal I}_{pq}(t-r)+
6\frac{{\cal{I}}_{pq}(t-r)}{r}+
15\frac{^{(-1)}{\cal{I}}_{pq}(t-r)}{r^2}+
15\frac{^{(-2)}{\cal{I}}_{pq}(t-r)}{r^3}\right]
\frac{\xi^i k^p x^q_N }{r^4}\\\nonumber\\\nonumber\mbox{}&&\hspace{-3 cm}
+4\frac{\tau}{r}\left[\dot{\cal I}_{pq}(t-r)+
\frac{15}{2}\frac{{\cal{I}}_{pq}(t-r)}{r}+
\frac{45}{2}\frac{^{(-1)}{\cal{I}}_{pq}(t-r)}{r^2}+
\frac{45}{2}\frac{^{(-2)}{\cal{I}}_{pq}(t-r)}{r^3}\right]
\frac{\xi^i x^p_N x^q_N }{r^5}\\\nonumber\\\nonumber\mbox{}&&\hspace{-2 cm}
+\frac{y}{r}\left[\ddot{\cal I}_{pq}(t-r)+
6\frac{\dot{\cal{I}}_{pq}(t-r)}{r}+
15\frac{{\cal{I}}_{pq}(t-r)}{r^2}+
15\frac{^{(-1)}{\cal{I}}_{pq}(t-r)}{r^3}\right]\frac{\xi^i x^p_N x^q_N
}{r^4}\;.\\\nonumber
\end{eqnarray}

We note that the angular displacement in astrometric positions of sources of light 
in the sky depends not only on 
quantities that are instantaneous functions of
retarded time, but also on integrals over 
time $^{(-1)}{\cal{I}}_{pq}(t-r)$ and 
$^{(-2)}{\cal{I}}_{pq}(t-r)$, which describe the 
whole past history of the source
of gravitational waves up to the moment of observation. Under usual
circumstances the influence of such integrals on the deflection of light is expected to
be small. However, this question deserves more detailed discussion and will be
discussed elsewhere.
\section{Discussion}

It is remarkable that among all the integrals 
required for  calculation of the
trajectory of the light ray, only 
$B_{ij}(\tau, {\bm{\xi}})$ enters
the expressions
(\ref{quad1}), (\ref{dyw}) for time delay and deflection angle.
Furthermore, it is remarkable that 
we need not know this integral explicitly, but only its
second and third
derivatives  with respect to 
impact parameter. 
These are given in equations (\ref{pzka})
and (\ref{pzkab}). With the knowledge of these derivatives, and 
derivatives of
other functions given in the previous section, we have 
complete
information about the functional structure of 
the relativistic time delay and the angle of light deflection 
produced by any localized gravitating system possessing a
time-dependent quadrupole moment ${\cal I}_{ij}(t)$. 

This structure indicates that the 
explicit time dependence of the quadrupole moment completely
determines the results of astrometric and timing observations.
We shall not consider
this problem in the present paper, leaving it for future exploration.

Our concern in this section is the simplification of the 
general formalism
developed in the foregoing text. In order to do this 
we consider
three limiting cases:

\begin{enumerate}

\item The impact parameter $d$
is much smaller than  the distance from the localized source of 
gravitational waves to both the observer, $r$, and 
to the source of light, $r_0$.
The source of light is behind the source of gravitational waves
(see Figure \ref{smallimp1});

\item The impact parameter $d$
is much smaller than  the distance from the localized source of 
gravitational waves to the observer, $r$, and 
to the source of light, $r_0$.
The source of light is on the same side of the 
source of gravitational waves as the observer (see Figure \ref{smallimp2});

\item The distance $R$ from the source of light rays to the 
observer is much
smaller than distances from  the observer or from the 
source of light to the 
localized
source of gravitational waves. The impact parameter $d$ may be 
comparable with the 
distance from the deflector to observer or the source of light
(see Figure \ref{largeimp}).
\end{enumerate}
We will conventionally refer to the cases 1 and 2 as those of small impact 
parameter, with numerical values of $\tau_0<0$ and $\tau_0>0$ respectively. 
Case 3 is that of large impact parameter, and 
its small numerical
values are covered by the formalism as well, as will be clear  in
section 7.3 below.

\subsection {Case 1. Small Impact Parameter ($\tau_0<0$)}

\subsubsection{Asymptotic expansions of independent variables}

We shall assume in this section that the condition $d\ll{\rm min}[r,r_0]$ 
holds. Let $L={\rm min}[r,r_0]$ and recall that 
$\tau=\sqrt{r^2-d^2}$ and $\tau_0=-\sqrt{r_0^2-d^2}<0$ (see Figure
\ref{smallimp1}).  This yields
\vspace{0.3 cm}
\begin{equation}
\label{gfr}
y=\sqrt{r^2-d^2}-r=-\frac{d^2}{2r}-\frac{d^4}{8r^3}+...,
\end{equation}
\vspace{0.3 cm}
and
\vspace{0.3 cm}
\begin{equation}
\label{pxl}
y_0=-\sqrt{r_0^2-d^2}-r_0=-2r_0+\frac{d^2}{2r_0}+\frac{d^4}{8r_0^3}+...\;,
\\\nonumber
\end{equation}
where dots denote terms of higher order, $r$ is the constant
distance from the deflector to observer, and $r_0$ is the constant distance
from the deflector to the point of emission of light. Using these expansions we
find:
\vspace{0.3 cm}
\begin{equation}
\label{tnx}
t=t^{\ast}+r-\frac{d^2}{2r}+...,\hspace{1.5 cm}
t_0=t^{\ast}-r_0+\frac{d^2}{2r_0}+...\;.
\end{equation}
These can be used for  Taylor expansion of functions 
about the time $t^\ast$,
the moment
of the closest approach of light ray to the deflector. 
Specifically, if we assume convergence of this Taylor series 
we find:
\vspace{0.3 cm}
\begin{equation}
\label{tfl}
{\cal I}_{ij}(t-r)={\cal I}_{ij}(t^{\ast})-\frac{d^2}{2r}
\dot{\cal I}_{ij}(t^{\ast})+...\;,
\end{equation}
\vspace{0.3 cm}
\begin{equation}
\label{tfz}
{\cal I}_{ij}(t_0-r_0)={\cal I}_{ij}(t^{\ast}-2r_0)+\frac{d^2}{2r_0}
\dot{\cal I}_{ij}(t^{\ast}-2r_0)+...\;,
\end{equation}
where dots again denote terms of higher order. Convergence 
of the
time series given above requires:
\begin{equation}\label{requir}
\frac{\omega d^2}{c\;r}\ll 1\;, \quad\quad\mbox{and}\quad\quad\frac{\omega d^2}
{c\;r_0}\ll 1\;,
\end{equation}
where $\omega$ is the highest frequency of gravitational waves emitted
by the deflector. If the source of light rays and observer are at 
infinite distances from the deflector then the requirements (\ref{requir}) are
satisfied automatically, irrespective of the structure of the 
Fourier spectrum of the
quadrupole moment of the deflector. In 
practical situations such an assumption may not be always satisfied. For this
reason, it will be more natural to avoid the Taylor expansions of the
quadrupole moment with respect to retarded time. 
It is also worth noting that in the case of small impact parameter we have
\vspace{0.3 cm}
\begin{equation}
\label{gfrs}
\left(yr\right)^{-1}=-\frac{2}{d^2}+\frac{1}{2r^2}+\frac{d^2}{8r^4}+...,
\end{equation}
\vspace{0.3 cm}
and
\vspace{0.3 cm}
\begin{equation}
\label{pxls}
\left(y_0 r_0\right)^{-1}=-\frac{1}{2r_0^2}-\frac{d^2}{8r_0^4}+...\hspace{0.5
cm}.
\end{equation}
\vspace{0.3 cm}
The foregoing expansions then yield\vspace{0.3 cm}
\begin{eqnarray}
\label{qqq}
{\hat{\partial}}_{j}B_{pq}(\tau, {\bm {\xi}})&=&\left(
-2{\hat{\partial}}_{j}\ln d+\frac{\xi^j}{2r^2}\right){\cal I}_{jk}(t-r)+...\;,\\
\nonumber\\\label{qq1}
{\hat{\partial}}_{j}B_{pq}(\tau_0, {\bm {\xi}})&=&-\frac{\xi^j}{2r_0^2}
{\cal I}_{jk}(t_0-r_0)+...\;,\\\nonumber\\
\label{acu}
{\hat{\partial}}_{jk}B_{pq}(\tau, {\bm {\xi}})&=&
-2{\hat{\partial}}_{jk}\ln d\;{\cal I}_{pq}(t-r)+\frac{2}{r}n_j n_k
\dot{\cal I}_{pq}(t-r)+...\;,\\\nonumber\\
\label{acp}
{\hat{\partial}}_{jk}B_{pq}(\tau_0, {\bm {\xi}})&=&-\frac{1}{2r_0^2}P_{jk}
{\cal I}_{pq}(t_0-r_0)+...\;,\\\nonumber\\\label{asm}
{\hat{\partial}}_{ijk}B_{pq}(\tau, {\bm {\xi}})&=&
-2\left[{\cal I}_{pq}(t-r)+\frac{d^2}{2r}\dot{\cal I}_{pq}(t-r)\right]
{\hat{\partial}}_{ijk}\ln d+...\;,\\\nonumber\\\label{asop}
{\hat{\partial}}_{ijk}B_{pq}(\tau_0, {\bm
{\xi}})&=&O\left(\frac{1}{r_0^3}\right)\;,\\\nonumber
\end{eqnarray}
and
\begin{eqnarray}
\label{derd}
{\hat{\partial}}_{ijk}D_{pq}(\tau, {\bm {\xi}})&=&
-2r{\cal I}_{pq}(t-r){\hat{\partial}}_{ijk}\ln d-\frac{4n^i n^j n^k}{d}
{\dot{\cal I}}_{pq}(t-r)+...\;,\\\nonumber\\
\label{derivd}
{\hat{\partial}}_{ijk}D_{pq}(\tau_0, {\bm {\xi}})&=&
O\left(\frac{1}{r_0^3}\right)\;.\\\nonumber
\end{eqnarray}
In addition we have
\begin{eqnarray}
\label{uit}
\delta_Q(\tau,{\bm{\xi}}) &=&
\frac{1}{r}k^pk^q\dot{\cal{I}}_{pq}(t-r)+...\;,\\
\nonumber\\\label{mas}
\delta_Q(\tau_0,{\bm{\xi}}) &=&
O\left(\frac{1}{r_0^2}\right)\;.
\end{eqnarray}
We note that the leading terms of the expansions decay much faster 
(at least as $1/r_0^2$) at
the point of emission of light than those at the point of
observation. This indicates that the main contribution to the effects of time
delay and deflection of light arise along the path of the
light ray from the 
localized
source of gravitational waves to the observer. We discuss this question in more
detail in the following section.
 
The asymptotic expansions of integrals (\ref{31}) - (\ref{32}) describing
propagation of light rays in the static part of gravitational field of the
deflector are:
\vspace{0.3 cm}
\begin{equation}
\label{alt}
A(\tau,{\bm{\xi}})=-2\ln d+\ln 2r-\frac{d^2}{4r^2}+
...\hspace{0.5 cm},
\end{equation}
\vspace{0.3 cm}
\begin{equation}
\label{altov}
A(\tau_0,{\bm{\xi}})=-\ln 2r_0+\frac{d^2}{4r_0^2}+...\hspace{0.5 cm},
\end{equation}
\vspace{0.3 cm}
\begin{eqnarray}
\label{blat}
B(\tau,{\bm{\xi}})&=&-r-2r \ln d+r \ln 2r-\frac{d^2}{2r}\left[\frac{1}{2}-
\ln\left(\frac{d^2}{2r}
\right)\right]...\;,\\\nonumber\\\nonumber\\
\label{gnus}
B(\tau_0,{\bm{\xi}})&=&-r_0+r_0 \ln 2r_0-\frac{d^2}{2r_0}
\left(\frac{1}{2}+\ln 2 r_0\right)+...\;.\\\nonumber
\end{eqnarray}
These expansions are used for calculation of asymptotic expressions
for time delay and the angle of deflection of light rays.

\subsubsection{Asymptotic expressions for time delay and the angle of 
light deflection}

The static part of time delay and deflection angle are:
\vspace{0.3 cm}
\begin{eqnarray}
\label{mass}
\Delta_M&=&-4{\cal M} \ln d+2{\cal M} \ln (4r r_0)+...\hspace{0.5
cm},\\\nonumber\\
\label{spin}
\Delta_S&=&-4\epsilon_{jip}k^j{\cal S}^p {\hat{\partial}}_i 
\left[\ln d-\frac{1}{2}\ln (4r r_0)\right]+...\hspace{0.5 cm},\\\nonumber\\
\label{ma}
\alpha_M^i(\tau,{\bm{\xi}})&=&
4{\cal M}{\hat{\partial}}_i\left[ \ln d-\frac{1}{2}\ln (4r r_0)\right]
+...\hspace{0.5 cm},\\\nonumber\\
\label{sp}
\alpha_S^i(\tau,{\bm{\xi}})
&=&4\epsilon_{jpq}k^p{\cal S}^q {\hat{\partial}}_{ij} 
\left[\ln d-\frac{1}{2}\ln (4r r_0)\right]+...\hspace{0.5 cm},\\\nonumber\\
\label{mam}
\beta_M^i(\tau,{\bm{\xi}})&=&-\frac{r}{R}\alpha_M^i(\tau,{\bm{\xi}})+...\;,
\\\nonumber\\
\label{spm}
\beta_S^i(\tau,{\bm{\xi}})&=&-\frac{r}{R}\alpha_S^i(\tau,{\bm{\xi}})-
\frac{4}{R}P^{ij}{\cal S}^k\epsilon_{jkq}{\hat{\partial}}_{q}\ln d+...
\;,
\\\nonumber
\end{eqnarray}
where we have neglected the angle $\gamma^i(\tau,{\bm{\xi}})$ 
because it is small 
(recall that $\gamma^i(\tau,{\bm{\xi}})\simeq P^{ij}k^q
h^{TT}_{jq}$).

Asymptotic expressions for the time delay and angle of deflection 
caused by the quadrupole moment are:
\vspace{0.3 cm}
\begin{eqnarray}
\label{quad}
\Delta_Q&=&-2{\cal I}_{ij}(t-r){\hat{\partial}}_{ij}\ln d+
\frac{1}{r}\left(2n_i n_j+k_i k_j\right)\dot{\cal I}_{ij}(t-r)
+...\;,\\\nonumber 
\end{eqnarray}
and
\vspace{0.3 cm}
\begin{eqnarray}
\label{angle}
\alpha_Q^i(\tau,{\bm{\xi}})&=&
2\left[{\cal I}_{jk}(t-r)+\frac{d^2}{2r}\dot{\cal I}_{jk}(t-r)\right]
{\hat{\partial}}_{ijk}\ln d 
+...\hspace{0.5 cm},\\\nonumber\\
\label{betanagle}
\beta_Q^i(\tau,{\bm{\xi}})&=&-\frac{r}{R}\alpha_Q^i(\tau,{\bm{\xi}})
-\frac{4}{R}\left[k^j{\cal I}_{jk}(t-r){\hat{\partial}}_{ik}\ln d+\frac{1}{2}
\xi^i {\dot{\cal I}}_{jk}(t-r){\hat{\partial}}_{jk}\ln d\right]+...\;,
\end{eqnarray}
where $n^i=\xi^i/d$ is the unit vector directed along the impact parameter, 
$R=|{\bf x}-{\bf x}_0|$, and dots denote 
terms of higher order \cite{53}.
The angle $\beta^i(\tau_0,{\bm{\xi}})$ at the point of emission of light is
negligibly small and, for this reason, its exact expression has been not shown. 

Our calculations show that the time dependent part of the time delay and 
light deflection by the quadrupole moment of a
localized source of gravitational field fall off in the first
approximation 
as the \underline{\it inverse square} and \underline{\it inverse cube} of
the impact parameter $d$ respectively. For this reason there is no
magnification of the gravitational wave signal in astrometric 
or pulsar timing
observations as some authors have suggested \cite{19} - \cite{21}. 
In particular, 
terms proportional  to
$1/d$, or even to $1/d^2$
, appear only in terms of high 
order  in the expansion (\ref{angle}) and are always multiplied by
the factor $1/r$ to some power. 

The first term of formula (\ref{quad}) was first derived by Sazhin
\cite{7} for the special case of 
a binary system with a specific orientation of its
orbital plane. 
Our derivation of formula (\ref{angle}) 
improves and gives independent confirmation 
of the result established previously by Damour \& Esposito-Far\`{e}se \cite{23} 
using another mathematical technique based on application of 
Fourier transform and pure harmonic coordinates. 
For completeness we have repeated the calculations of 
Damour \& Esposito-Far\`{e}se \cite{23} for 
the effect of deflection of light rays by 
localized sources of gravitational waves in ADM rather than
harmonic coordinates (see Appendix A). 
The result 
coincides completely with that of Damour \& Esposito-Far\`{e}se \cite{23}
and clearly demonstrates the gauge invariance of the result.
However, our technique is more general and powerful.
Our formalism is valid
for any relative position of observer,
source of light, and source of gravitational waves,
and with finite or infinite separations.
The method developed by Damour \& Esposito-Far\`{e}se \cite{23} is
valid only for infinite separations and for
small values of impact parameter. 
In particular, we note that while 
Damour \& Esposito-Far\`{e}se \cite{23} find that the deflection depends on the
time $t^{\ast}$ 
of the closest approach of light to
the deflector, 
our calculation shows that it depends on the retarded
time $t-r$. This difference is insignificant for extremely large separation of
the
light source and observer from the deflector,
and small impact parameter, but it
can be important in the cases of finite distances
or large impact parameter.

It is important to realize that in the case of a small impact parameter, 
the basic time-dependent contribution to the bending of light and time 
delay by the gravitational field of a localized source of gravitational waves 
comes from the static part of the near-zone gravitational field of 
the source taken at the retarded  time (cf. formulae (50)-(53) 
from \cite{24}). 
In this respect it is worth emphasizing that the formula for 
the bending of light given in paper \cite{23} as well as in Appendix A 
is valid under two assumptions: 1) the impact parameter $d$ is small 
compared with the distance to the observer $r$, 2) the velocity of matter 
inside the source of gravitational radiation is much smaller than the speed of 
light (the slow-motion approximation).

The first assumption is rather trivial, since the impact parameter $d$ is the 
only finite distance when the source of light and observer are at infinity.
The second assumption appears because paper \cite{23} uses the 
Taylor 
expansion of the Fourier image of the tensor of energy-momentum of matter with 
respect to wave vector ${\bf k}$ (see equations (3.3) and (3.4) of paper 
\cite{23}). This expansion is 
mathematically equivalent to the use of a slow-motion approximation \cite{53a}
which, in particular, restricts the nature of the source of gravitational 
waves so that its Fourier spectrum is not allowed to include too high 
frequencies.

In contrast, the general formalism given in the present paper produces results 
(\ref{quad}) and (\ref{angle})  applicable to arbitrary sources of
gravitational waves, including gravitational radiation bursts with  internal 
velocity of matter comparable to the speed of light \cite{53b}. 
Moreover, we do not assume positions of observer and the source of light to be 
at infinity. 
 
If we introduce the notion of the transverse-traceless (TT) and longitudinal
(L) tensors \cite{24}, \cite{33} with respect to the direction of propagation of light rays 
\begin{eqnarray}\label{TTT}
{\cal I}_{ij}^{TT}&=&{\cal I}_{ij}+
\frac{1}{2}\left(\delta_{ij}+k_i k_j\right)k_p k_q\; {\cal
I}_{pq}-\left(\delta_{ip}k_j k_q+\delta_{jp}k_i k_q\right)\;{\cal
I}_{pq}\;,\\\nonumber\\\label{long}
{\cal I}_{ij}^{L}&=&k_i k_p{\cal I}_{jp}+k_j k_p{\cal I}_{ip}-k_i k_j\left(k_p
k_q {\cal I}_{pq}\right)\;,\\\nonumber 
\end{eqnarray}
the expressions (\ref{quad})-(\ref{angle}) are reduced to the form
\begin{eqnarray}
\label{quadTT}
\Delta_Q&=&-2{\cal I}_{ij}^{TT}(t-r)\;{\hat{\partial}}_{ij}\ln d+
\frac{2}{r}n^i n^j\;\dot{\cal I}_{ij}^{TT}(t-r)
+...\;,\\\nonumber\\ \label{angleTT}
\alpha_Q^i(\tau,{\bm{\xi}})&=&
2\left[{\cal I}_{jk}^{TT}(t-r)+\frac{d^2}{2r}\;\dot{\cal I}_{jk}^{TT}(t-r)\right]
{\hat{\partial}}_{ijk}\ln d 
+...\hspace{0.5 cm},\\\nonumber\\
\label{b}
\beta_Q^i(\tau,{\bm{\xi}})&=&-\frac{r}{R}\alpha_Q^i(\tau,{\bm{\xi}})
-\frac{4}{R}\left[k^j{\cal I}_{jk}^{L}(t-r){\hat{\partial}}_{ik}\ln
d+\frac{1}{2}\;
\xi^i {\dot{\cal I}}_{jk}^{TT}(t-r){\hat{\partial}}_{jk}\ln d\right]+...\;,
\\\nonumber
\end{eqnarray}
which reveals explicitly that only the transverse-traceless part of the 
quadrupole
moment of the localized source of gravitational waves contributes 
to the leading terms. However, terms of higher
order  can depend on the longitudinal 
part of the quadrupole moment as
well.

It is interesting to see that if we apply the expansions (\ref{tfl})-(\ref{tfz}),
use the approximation of a 
gravitational lens, and omit all terms depending on
time derivatives of the quadrupole moment,
the expressions for the time delay and the angle of light deflection can be
reduced to the formulae \cite{54}
\begin{eqnarray}
\label{timed}
t-t_0&=&|{\bf x}-{\bf x}_0|-
4\psi+2{\cal M}\ln(4r r_0)\;,\hspace{2 cm}\alpha_i=4{\hat{\partial}}_i\psi\;,\\\nonumber
\end{eqnarray}
where $\psi$ is the gravitational lens potential \cite{55}
having the form 
\vspace{0.3 cm}
\begin{eqnarray}
\label{damour}
\psi&=&\left[{\cal M}+\epsilon_{jpq} k^p{\cal S}^q{\hat{\partial}}_j+
\frac{1}{2}\;{\cal I}_{pq}^{TT}(t^{\ast})\;{\hat{\partial}}_{pq}
\right]\ln d\;.
\end{eqnarray}
Scrutiny of the multipole structure 
of $\psi$ in cosmological gravitational lenses 
may reveal the
presence of dark matter in the lens and identify the position of its center of 
mass, velocity and density distribution.

Expression (\ref{damour}) includes explicit dependence on mass, spin, and
quadrupole moment of the deflector and generalizes that given by Damour \&
Esposito-Far\`{e}se \cite{23} by
accounting for the spin multipole. A similar result 
for the gravitational lens
potential was obtained independently by Kopeikin \cite{24} in the case of
a
stationary gravitational field for the deflector. 
The fact that the deflection angle can be represented as a gradient of the 
gravitational lens potential $\psi$ explicitly indicates that  
the, so-called, frame-dragging effect in gravitational 
lenses \cite{56} can give 
a noticeable contribution to  
the deflection angle. 
Frame-dragging also produces a
small displacement of the image of the background 
source from the plane formed by the two vectors 
directed from the observer 
toward the image of the light source and toward the 
gravitational lens. This torsional displacement of the image is 
produced only by the component of spin of the deflector directed along the light 
ray (see the second term in equation (\ref{spm}). The overall effect of the torsion is of 
order $d/r$ smaller than the main terms in the expression
(\ref{damour}). These remarks 
dispel a seemingly common opinion that rotation of the
deflector prevents representation of the deflection angle as a gradient of a
gravitational lens potential.
Similar conclusions can be derived from 
\cite{24} and \cite{32}.  
Ib\'a\~nez \& Martin \cite{57} and Ib\'a\~nez \cite{58} give a formula 
for effects of frame-dragging equivalent to the spin-dependent term 
in (\ref{damour}), although they do not calculate all necessary 
integrals or estimate residual terms.

Taking into account formula (\ref{dop}) and expressions for $\alpha^i$, 
$\beta^i$, and $\gamma$ we obtain the 
vector equation for a gravitational lens
\begin{eqnarray}
\label{lens}
s^i
&=&K^i+\frac{r_0}{R}\;\alpha^i\;,
\\\nonumber
\end{eqnarray}
where $\alpha^i$ is given by relationships (\ref{timed}), (\ref{damour}) 
and we have taken into account that in the case under consideration $R\simeq 
r+r_0$. One recognizes that when distances are finite the deflection angle 
with respect to vector $K^i$ is not simply $\alpha^i$ but the product of 
$r_0/R$ and $\alpha^i$. In the limit when $K^i\rightarrow k^i$, which is
equaivalent to $\beta^i\rightarrow 0$, or $r={\rm const.}$, $r_0\rightarrow
\infty$ the observed 
angle of deflection approaches the total angle of deflection 
$\alpha^i$, as it must in this limiting case. 

\subsection {Case 2. Small Impact Parameter ($\tau_0>0$)}

\subsubsection{Asymptotic expansions of independent variables}

We shall again assume in this section that the condition 
$d\ll{\rm min}[r,r_0]$ holds and that 
$\tau=\sqrt{r^2-d^2}$ and $\tau_0=\sqrt{r_0^2-d^2}>0$ (see Figure
\ref{smallimp2}).  This yields
\vspace{0.3 cm}
\begin{equation}
\label{gfr1}
y=\sqrt{r^2-d^2}-r=-\frac{d^2}{2r}-\frac{d^4}{8r^3}+...,
\end{equation}
\vspace{0.3 cm}
and
\vspace{0.3 cm}
\begin{equation}
\label{pxl1}
y_0=\sqrt{r_0^2-d^2}-r_0=-\frac{d^2}{2r_0}-\frac{d^4}{8r^3_0}+...\;,
\\\nonumber
\end{equation}
where dots denote terms of higher order, $r$ is the constant
distance from the deflector to observer, and $r_0$ is the constant distance
from the deflector to the point of emission of light. Using these expansions we
obtain the following decompositions
\vspace{0.3 cm}
\begin{equation}
\label{tnx1}
t=t^{\ast}+r-\frac{d^2}{2r}+...,\hspace{1.5 cm}
t_0=t^{\ast}+r_0-\frac{d^2}{2r_0}+...\;.
\end{equation}
These can be used for 
Taylor expansion of functions that depend on 
retarded time about the time $t^\ast$. 
In this case $t^{\ast}$ is
the moment
of  closest approach of the light ray 
trajectory extrapolated backward 
to the deflector (see Figure \ref{smallimp2}). 
If we assume convergence of this Taylor series 
we find:
\vspace{0.3 cm}
\begin{equation}
\label{tfl1}
{\cal I}_{ij}(t-r)={\cal I}_{ij}(t^{\ast})-\frac{d^2}{2r}
\dot{\cal I}_{ij}(t^{\ast})+...\;,
\end{equation}
\vspace{0.3 cm}
\begin{equation}
\label{tfz1}
{\cal I}_{ij}(t_0-r_0)={\cal I}_{ij}(t^{\ast})-\frac{d^2}{2r_0}
\dot{\cal I}_{ij}(t^{\ast})+...\;,
\end{equation}
where dots again denote terms of higher order. 
We also have
\vspace{0.3 cm}
\begin{equation}
\label{gfrs1}
\left(yr\right)^{-1}=-\frac{2}{d^2}+\frac{1}{2r^2}+\frac{d^2}{8r^4}+...,
\end{equation}
\vspace{0.3 cm}
and
\vspace{0.3 cm}
\begin{equation}
\label{pxls1}
\left(y_0 r_0\right)^{-1}=-\frac{2}{d^2}+\frac{1}{2r_0^2}+\frac{d^2}{8r_0^4}+...\hspace{0.5
cm}.
\end{equation}
\vspace{0.3 cm}
The foregoing expansions yield\vspace{0.3 cm}
\begin{eqnarray}
\label{qqq1}
{\hat{\partial}}_{j}B_{pq}(\tau, {\bm {\xi}})&=&\left(
-2{\hat{\partial}}_{j}\ln d+\frac{\xi^j}{2r^2}\right){\cal I}_{jk}(t-r)+...\;,\\
\nonumber\\\label{qq11}
{\hat{\partial}}_{j}B_{pq}(\tau_0, {\bm {\xi}})&=&\left(
-2{\hat{\partial}}_{j}\ln d+\frac{\xi^j}{2r_0^2}\right){\cal I}_{jk}(t_0-r_0)
+...\;,\\\nonumber\\
\label{acu1}
{\hat{\partial}}_{jk}B_{jk}(\tau, {\bm {\xi}})&=&
-2{\hat{\partial}}_{jk}\ln d\;{\cal I}_{jk}(t-r)+\frac{2}{r}n_j n_k
\dot{\cal I}_{jk}(t-r)+...\;,\\\nonumber\\
\label{acp1}
{\hat{\partial}}_{jk}B_{jk}(\tau_0, {\bm {\xi}})&=&
-2{\hat{\partial}}_{jk}\ln d\;{\cal I}_{jk}(t_0-r_0)+\frac{2}{r_0}n_j n_k
\dot{\cal I}_{jk}(t_0-r_0)+...\;,\\\nonumber\\
\label{asm1}
{\hat{\partial}}_{ijk}B_{jk}(\tau, {\bm {\xi}})&=&
-2\left[{\cal I}_{jk}(t-r)+\frac{d^2}{2r}\dot{\cal I}_{jk}(t-r)\right]
{\hat{\partial}}_{ijk}\ln d+...\;,\\\nonumber\\
\label{asop1}
{\hat{\partial}}_{ijk}B_{jk}(\tau_0, {\bm
{\xi}})&=&
-2\left[{\cal I}_{jk}(t_0-r_0)+\frac{d^2}{2r_0}\dot{\cal I}_{jk}(t_0-r_0)\right]
{\hat{\partial}}_{ijk}\ln d+...
\;.\\\nonumber
\end{eqnarray}
and
\begin{eqnarray}
\label{derd1}
{\hat{\partial}}_{ijk}D_{pq}(\tau, {\bm {\xi}})&=&
-2r{\cal I}_{pq}(t-r){\hat{\partial}}_{ijk}\ln d-\frac{4n^i n^j n^k}{d}
{\dot{\cal I}}_{pq}(t-r)+...\;,\\\nonumber\\
\label{derivd1}
{\hat{\partial}}_{ijk}D_{pq}(\tau_0, {\bm {\xi}})&=&
-2r_0{\cal I}_{pq}(t_0-r_0){\hat{\partial}}_{ijk}\ln d-\frac{4n^i n^j n^k}{d}
{\dot{\cal I}}_{pq}(t_0-r_0)+...\;.\\\nonumber
\end{eqnarray}
In addition we have
\begin{eqnarray}
\label{uit1}
\delta_Q(\tau,{\bm{\xi}}) &=&
\frac{1}{r}k^pk^q\dot{\cal{I}}_{pq}(t-r)+...\;,\\
\nonumber\\
\label{mas1}
\delta_Q(\tau_0,{\bm{\xi}}) &=&
\frac{1}{r_0}k^pk^q\dot{\cal{I}}_{pq}(t_0-r_0)+...\;.
\end{eqnarray}
We note that the leading terms of the expansions now have the same dependence
on the distance of the point of emission of light 
and of the point of
observation from the source of gravitational waves. 
If the source of light is closer to the source of gravitational waves
than the observer,
it makes the largest contribution to the effects of time
delay and deflection of light. 
 
The asymptotic expansions of integrals (\ref{31}) - (\ref{32}) describing
propagation of light rays in the static part of the gravitational field of the
deflector are:
\vspace{0.3 cm}
\begin{equation}
\label{alt1}
A(\tau,{\bm{\xi}})=-2\ln d+\ln (2r)-\frac{d^2}{4r^2}+
...\hspace{0.5 cm},
\end{equation}
\vspace{0.3 cm}
\begin{equation}
\label{altov1}
A(\tau_0,{\bm{\xi}})=-2\ln d+\ln (2r_0)-\frac{d^2}{4r_0^2}+...\hspace{0.5 cm},
\end{equation}
\vspace{0.3 cm}
\begin{eqnarray}
\label{blat1}
B(\tau,{\bm{\xi}})&=&-r-2r \ln d+r \ln (2r)-\frac{d^2}{2r}\left[\frac{1}{2}-
\ln\left(\frac{d^2}{2r}
\right)\right]...\;,\\\nonumber\\\nonumber\\
\label{gnus1}
B(\tau_0,{\bm{\xi}})&=&-r_0-2r_0 \ln d+r_0 \ln (2r_0)-\frac{d^2}{2r_0}
\left[\frac{1}{2}-
\ln\left(\frac{d^2}{2r_0}\right)\right]+...\;.\\\nonumber
\end{eqnarray}
These expansions are used for calculation of asymptotic expressions
for time delay and the angle of deflection of light rays.

\subsubsection{Asymptotic expressions for time delay and the angle of 
light deflection}

The static part of time delay and deflection angle are given by:
\vspace{0.3 cm}
\begin{eqnarray}
\label{mass11}
\Delta_M=2{\cal M} \left[\ln \left(\frac{r}{r_0}\right)+
\frac{d^2}{4}\left(\frac{1}{r_0^2}-\frac{1}{r^2}\right)\right]
+...\hspace{0.5 cm},
\end{eqnarray} 
\vspace{0.3 cm}
\begin{eqnarray}
\label{spin11}
\Delta_S&=&\epsilon_{ijp}k^j{\cal S}^p\xi^i\left(
\frac{1}{r_0^2}-\frac{1}{r^2}\right) +...\hspace{0.5 cm},\\\nonumber
\end{eqnarray}
Expressions for $\alpha_M^i$, $\alpha_S^i$, and $\alpha_Q^i$ will be the same
as in equations (\ref{ma}), (\ref{sp}), and (\ref{angle}) because they are taken 
at the point of observation only. Expressions for $\beta_M^i$, $\beta_S^i$, and
$\beta_Q^i$ are given at the point of observation by equations
(\ref{mam}), (\ref{spm}), and (\ref{betanagle}). 
Expressions for $\beta_M^i$, $\beta_S^i$, and
$\beta_Q^i$ at the point of emission of light are given by the same equations
(\ref{mam}), (\ref{spm}), and (\ref{betanagle}) after 
substituting $r_0$ for $r$. The relativistic 
perturbation $\gamma^i$ is calculated 
in
equation (\ref{gamma}).

The asymptotic expression for the time delay  
caused by the quadrupole moment is:
\vspace{0.3 cm}
\begin{eqnarray}
\label{quad11}
\Delta_Q&=&-2\left[{\cal I}_{ij}(t-r)-{\cal I}_{ij}(t_0-r_0)\right]
{\hat{\partial}}_{ij}\ln d\\\nonumber\\\mbox{}&&
+
\left(2n_i n_j+k_i k_j\right)\left[\frac{\dot{\cal I}_{ij}(t-r)}{r}-
\frac{\dot{\cal I}_{ij}(t_0-r_0)}{r_0}\right]+...\;. 
\end{eqnarray}
One might think that the effect of retardation is again inversely
proportional
to the 
square of impact parameter $d$. However, this  is actually true
only for sources of gravitational waves with rapidly varying quadrupole moment.
If motion of matter inside the localized source of gravitational waves is slow, then
conditions (\ref{requir}) apply. In this 
case, the real amplitude of the effect
becomes extremely small, being 
inversely proportional to $1/r^2$ and $1/r^2_0$.
   
The asymptotic
expression for the observed direction $s^i$ to the source of light is
derived from the basic formula (\ref{dop}) and is:
\begin{eqnarray}
\label{tauzer}
s^i&=&K^i-\frac{2r_0}{R}\left[{\cal I}_{jk}(t-r)-
{\cal I}_{jk}(t_0-r_0)\right]{\hat{\partial}}_{ijk}\ln d\\\nonumber\mbox{}&&
-\frac{4k^j}{R}\left[{\cal I}_{jk}(t-r)-{\cal I}_{jk}(t_0-r_0)\right]
{\hat{\partial}}_{ik}\ln d\\\nonumber\mbox{}&&
-\frac{2\xi^i}{R}\left[\dot{\cal I}_{jk}(t-r)-
\dot{\cal I}_{jk}(t_0-r_0)\right]{\hat{\partial}}_{jk}\ln
d+\gamma^i(\tau,{\bm{\xi}})+...\;,\\\nonumber
\end{eqnarray} 
where we have accounted  for the approximate equality 
$R\simeq r-r_0$ valid in the case of $\tau_0>0$. One can see that deflection 
angle is small in the expression given. Moreover, if
we again assume that motion of the matter is slow, 
then the observed deflection is
even smaller and is inversely proportional to $1/(rR)$ and $1/(r_0R)$.

\subsection{Case 3. Large Impact Parameter}
\subsubsection{Asymptotic expansions of independent variables}
In this limiting case we assume that the distance $R$ 
between observer and source
of light is much smaller than $r$ and $r_0$, their respective distances from 
the deflector (see Figure
\ref{largeimp}). Then we have
\begin{eqnarray}\label{larg}
r_0^2&=&r^2-2r R
\cos\theta+R^2=r^2\left(1-\frac{2R}{r}\cos\theta+\frac{r^2}{r^2}\right)\;,
\end{eqnarray}
which leads to the expansions\vspace{0.3 cm}
\begin{eqnarray}
\label{expan}
r_0&=&r-R\cos\theta+...\;,\\\nonumber\\\label{q1}
\frac{1}{r_0}&=&\frac{1}{r}\left(1+\frac{R}{r}\cos\theta\right)+...\;.
\end{eqnarray}
The time parameters are 
\begin{equation}
\label{free}
\tau=r\cos\theta\;,\quad\quad\mbox{and}\quad\quad \tau_0=\tau-R\;.
\end{equation} 
Their numerical values 
may be larger or less than zero. 
The following exact equalities hold:
\begin{eqnarray}\label{q2}
d&=&r\sin\theta\;,\\
y&=&\tau-r=r(\cos\theta-1)
\;,\\
\left(yr\right)^{-1}&=&\frac{1}{r^2(\cos\theta-1)}\;.
\end{eqnarray}
In addition, we have asymptotic expansions
\begin{eqnarray}
\label{q3}
y_0&=&\tau_0-r_0=y\left(1+\frac{R}{r}\right)+...\;,\\
\left(y_0r_0\right)^{-1}&=&\frac{1}{yr}+\frac{R}{r^3}+...\;,\\ \\
t_0-r_0&=&t-r+R(\cos\theta-1)+...\;.
\end{eqnarray}
Thus, we can decompose any function of the time argument $t_0-r_0$ in a
Taylor series
with respect to the retarded time $t-r$ if convergence is
assumed \cite{59}. For example,
\begin{eqnarray}\label{texp}
{\cal{I}}_{ij}(t_0-r_0)&=&{\cal{I}}_{ij}(t-r)+R(\cos\theta-1)\;
\dot{\cal{I}}_{ij}(t-r)+...\;.
\end{eqnarray}
Finally, we note that the vector $\xi^i$ corresponding to impact parameter $d$ 
can be
represented as
\begin{eqnarray}\label{vecxi}
\xi^i&=&r\left(N^i-k^i\cos\theta\right)\;,
\end{eqnarray}
where $N^i=-K_0^i=x^i/r$, $|{\bf N}|=1$, 
and $k^i$ is the unit vector in the direction from the source of light to
observer \cite{60}. 

\subsubsection{Asymptotic expressions for time delay and the angle of 
light deflection}

In this section all asymptotic expressions for relativistic effects 
are given only up to leading terms
of order $1/r$ and $1/r_0$. For this reason all residual terms of order $1/r^2$
and $1/r^2_0$
are omitted in subsequent formulae without note. 
Using asymptotic expansions of functions from the previous section and reducing
similar terms we obtain 
\vspace{0.3 cm}
\begin{equation}
\label{pas}
\Delta_Q=\frac{1}{1-\cos\theta}\left[k^i k^j-2k^i N^j\cos\theta+
\frac{1}{2}\left(1+\cos^2\theta\right)N^iN^j\right]
\biggl\{\frac{\dot{\cal I}_{ij}(t-r)}{r}-
\frac{\dot{\cal I}_{ij}(t_0-r_0)}{r_0}\biggr\}\;,\\\nonumber
\end{equation}
where $\cos\theta={\bf k}\cdot{\bf N}={\bf K}\cdot{\bf K}_0$ (see Figures
\ref{bundle} and \ref{largeimp}).
We note that the expression for time delay given above can be further
simplified if the definition of ``transverse-traceless" tensor
with respect to the direction $N^i$ is applied \cite{24},
\cite{33}:
\begin{eqnarray}\label{TT}
{\cal I}_{ij}^{TT}&=&{\cal I}_{ij}+
\frac{1}{2}\left(\delta_{ij}+N_i N_j\right)N_p N_q\; {\cal
I}_{pq}-\left(\delta_{ip}N_j N_q+\delta_{jp}N_i N_q\right)\;{\cal I}_{pq}\;,
\end{eqnarray}
where the projection is onto the plane orthogonal to unit vector $N^i$.
Formula (\ref{pas}) for time delay now assumes the form
\begin{eqnarray}\label{timedel}
\Delta_Q&=&\frac{k^i k^j}{1-\cos\theta}\left[
\frac{\dot{\cal I}_{ij}^{TT}(t-r)}{r}-
\frac{\dot{\cal I}_{ij}^{TT}(t_0-r_0)}{r_0}\right]\;.
\end{eqnarray}
Differentiation of $\Delta_Q$ with respect to time gives the frequency shift 
due to a remote localized source of gravitational waves
\begin{eqnarray}\label{freq}
z_g(t,t_0)&=&1-\frac{dt}{dt_0}=-\frac{1}{2}\;
\frac{k^i k^j}{1-{\bf k}\cdot{\bf N}}
\left[h_{ij}^{TT}(t-r)-h_{ij}^{TT}(t_0-r_0)\right]\;,
\end{eqnarray}
where the metric $h_{ij}^{TT}$ is defined by the equation (\ref{adm4}) and
taken in the leading order approximation with respect to $1/r$. We
recognize that the expression (\ref{freq}) is a generalization of the analogous
formula for $z_g$ obtained previously by Mashhoon \& Seitz \cite{61} 
in the case of a plane
gravitational wave. This exact coincidence demonstrates the power of our 
formalism, 
which both reproduces well-known results and yields new
observational 
predictions for relativistic effects in the propagation of light
rays in the field of an 
arbitrary source of gravitational waves \cite{62}.

Repeating the calculations for the angle of light deflection under the assumption that
the
wavelength, $\lambda$, of gravitational waves emitted by the localized source
is smaller than the distance $R$ between source of light and observer,
we come to the following result:
\begin{eqnarray}
\label{asdr}
\alpha_Q^i&=&\frac{1}{1-\cos\theta}\left[\left(\cos\theta-2\right)
\left(k^ik^pk^q+2 k^ik^pN^q\cos\theta\right)+
\left(\cos^2\theta-2\cos\theta-1\right)\times\right.
\\\nonumber\\\nonumber\mbox{}&&\left.\times
\left(\frac{1}{2}k^iN^pN^q\cos\theta-N^iN^pN^q\right)+
N^ik^pk^q-2N^iN^pk^q\right]
\biggl\{\frac{\ddot{\cal I}_{pq}(t-r)}{r}\biggl\}
\\\nonumber\\\nonumber\mbox{}&&
+2\left(k^p- N^p\cos\theta\right)
\biggl\{\frac{\ddot{\cal I}_{ip}(t-r)}{r}\biggl\}\;.\\\nonumber
\end{eqnarray}
Transformation of this result 
using relationship (\ref{TT}) and expression (\ref{adm4}) for
$h_{ij}^{TT}$, where only leading terms of order
$1/r$ are retained, reveals that\vspace{0.3 cm}   
\begin{eqnarray}
\label{klon}
\alpha_Q^i&=&\frac{1}{2}\;
\frac{k^p k^q }{1-{\bf k}\cdot{\bf N}}\left[
\left({\bf k}\cdot{\bf N}-2\right)k^i+N^i\right]h_{pq}^{TT}(t-r)
+k^p h_{ip}^{TT}(t-r)\;,\\\nonumber
\end{eqnarray}
and, because the vector $\beta^i$ is small,\vspace{0.5 cm}
\begin{eqnarray}
\label{ssa}
s^i&=&K^i+\alpha^i_Q+\gamma^i+...\;.\\\nonumber
\end{eqnarray}
where the ellipsis designates 
unimportant terms of higher order with
respect to $1/r$ \cite{63}, and we
have neglected the constant deflection caused by mass-monopole and spin-dipole 
dependent terms.
One sees again that only the transverse-traceless component $h_{ij}^{TT}$ 
of the metric tensor appears in the final expression. 

It is worthwhile to stress that the observed optical direction to the source of 
light given by the formula (\ref{ssa}) coincides with that which can be obtained
by means of VLBI observations. Indeed, it is easy to confirm that equation 
(\ref{ssa}) can be re-written as follows \cite{64}
\begin{eqnarray}
\label{direc}
s^i&=&K^i+\frac{1}{2}\;
\frac{K^i+N^i}{1+{\bf K}\cdot{\bf N}}\;K^p K^q h_{pq}^{TT}(t-r)
-\frac{1}{2}K^p h_{ip}^{TT}(t-r)\;.\\\nonumber
\end{eqnarray}
The direction to the source of electromagnetic waves measured by VLBI is determined
as difference between times of arrival of the wave to the first and second
antennas. Taking into account equations (\ref{qer}) and (\ref{timedel}) for the
first and second observing sites, and assuming that the time difference  
$t_2-t_1$ in observation of the radio signal at the observatories is small
compared to the period of gravitational waves, we find
\vspace{0.3 cm}
\begin{eqnarray}
\label{timedif}
t_2-t_1&=&-\left({\bf K}+\frac{1}{2}\;
\frac{K^i+N^i}{1+{\bf K}\cdot{\bf N}}\;K^p K^q h_{pq}^{TT}(t-r)
\right)
\cdot({\bf x}_2-{\bf x}_1)\;.
\end{eqnarray}
If the baseline vector measured in the local inertial frame is denoted as ${\bf
b}$ and the transformation (\ref{trans}) is taken into account,
\vspace{0.3 cm}
\begin{eqnarray}
\label{end}
x^i_2-x^i_1&=&b^i-\frac{1}{2}\;h^{TT}_{ij}(t-r)b^j+O({\bf b}^2)\;.
\end{eqnarray}
We confirm that
\vspace{0.3 cm}
\begin{eqnarray}
\label{finish}
t_2-t_1&=&-{\bf s}\cdot{\bf b}\;,
\end{eqnarray}
where the vector $s^i$ is given by formula (\ref{direc}), which proves our
statement. It is worth emphasizing that equation (\ref{direc}) was obtained
independently by Pyne {\it et al.} (\cite{16}, see formula (47)). Their formalism, however,
has a limited region of application. Extension of the formalism of 
Pyne {\it et al.} \cite{16} was one of the motivation of the present work. 

\section{Conclusions}

The most accurate astrometric measurements are differential.  They
measure the angle between 2 sources.  The highest accuracy is
attainable when the sources are close to each other in the sky.  In contrast, 
angular deflection by gravitational waves varies only over large angles in the
general case of large impact parameter. Specifically, in such a case
the bending angle depends only on the metric in the
neighborhood of the observer and its first derivatives, as in 
equations (\ref{klon}), (\ref{ssa}). 
It thus can vary only as a quadrupole and the derivative of a
quadrupole, over the sky.  Similarly, equations (\ref{dyw}),
(\ref{angle}) and (\ref{tauzer}) depend on the mass quadrupole moment
${\cal {I}}_{ij}$ and its first and second derivatives. Note that the 
the angle of light deflection (\ref{dyw}) 
involves the time integrals of ${\cal {I}}_{ij}(t-r)$ 
which may be interpreted as the presence of the ``kinematic resonance effect'' 
\cite{14}; however, this term is small, as
discussed above.  In the context of a purely locally-determined
deflection angle, it is not unexpected that lines of sight that pass
close to the deflector show almost purely the static effect, as was shown in
section 7.

The magnitudes of the leading terms in the limiting forms for the
deflection angle $\alpha_Q$, in equations (\ref{dyw}), (\ref{angle}) and
(\ref{tauzer}) are $\alpha_Q\sim \Omega^2 G M a^2/c^4 r$, where $M$ is
the mass of the deflector, and $a$ is its dimension.  The frequency of
the gravitational waves is $\Omega$. For a gravitationally bound binary 
system with a circular orbit, 
$\Omega$ is twice the orbital frequency \cite{65}.  We can use Kepler's third law to express
this in the form $\alpha_Q\sim \Omega^{2/3}(GM)^{5/3}/c^4 r$, or
alternatively, $\alpha_Q\sim 2.4\times 10^{-14}(M/M_{\odot})^{5/3}
P_{\rm sec}^{-2/3} (r_{\rm kpc})^{-1}\;{\rm arcsec}$ where $P_{\rm sec}$ is
the orbital period of the binary system.  For a contact
white-dwarf binary at 200~pc, the expected deflection is about
$7\times 10^{-13}$~{\rm arcsec}, with a period of about 1000~sec.  For
a supermassive black-hole binary, with mass $10^6~M_{\odot}$ and
period 10~yr at a distance of 1~Mpc, the expected deflection is about
$5\times 10^{-11}$~{\rm arcsec}.

Because the effect varies smoothly over the sky, the presently available
astrometric accuracies are a few microarcseconds.  Higher accuracies
are attainable only over smaller angles.  Very-long baseline
interferometry of a suite of radio sources attains 
microarcsecond accuracy, over
periods of days to years.  Specially-designed observations sensitive
to source motions of minutes or hours might attain higher accuracy,
perhaps as much as an order of magnitude better.  Clearly, detection
of deflection of light rays by gravitational waves
from nearby localized sources 
is not a goal for the near future because of its smallness.
However the background gravitational wave noise may be, perhaps, measured.

The near-perfect cancellation of the effect in General Relativity
suggests that deflection of light by gravitational waves could be a
test of that theory in radiative regime \cite{66a}.  In a theory that does not 
posess the
symmetries that cause the deflection to vanish, we can only guess the
resulting deflection.  Such a guess might multiply the
general-relativistic $\alpha^i_Q$ by 3 factors.  The first factor, of
$r/d$, reflects the amplitude of the gravitational wave at the point of
closest approach, rather than at the observer.  The second factor,
some function of the distance to the source measured in
gravitational-wave wavelengths, perhaps $\ln (r/\lambda)$ \cite{67}, reflects
the cumulative effect of bending along the line of sight.  The final
factor, unknown, reflects the coupling of the non-general-relativistic
part of the wave to the source and its effect on the light ray.  For a
source an arcsecond from the deflectors described above, the first 2
factors can increase the effect by several orders of magnitude.  Moreover,
if the effect is not local to the observer, differential astrometry
across small angles can detect it, so that greater accuracy is
attainable. Given sufficiently strong departures
from General Relativity, the effect might be detectable.

\acknowledgments
{We are greatful to V.B. Braginsky, M.V. Sazhin, and P. Schneider for 
valuable and stimulating discussions. 
S.M. Kopeikin is pleasured to acknowledge the hospitality 
of G. Neugebauer and G. Sch\"afer and
other members of the Institute for Theoretical Physics of the Friedrich
Schiller University of Jena. 

The US National Science Foundation supported parts of this work (AST-9731584).
This work was partially
supported by the Th\"uringer Ministerium f\"ur Wissenschaft, Forschung und 
Kultur grant No B501-96060 and by the Max-Planck-Gesellschaft grant No
02160-361-TG74.}

\appendix
\section{Comparison to the paper by Damour \& Esposito-Far\`ese}

In this Appendix we rederive the results of the paper by
Damour \& Esposito-Far\`ese \cite{23} applying the generalized isotropic 
ADM coordinate conditions (For an application of the conditions in
post-Newtonian calculations, see e.g. \cite{41}).
This explicitly shows that the asymptotic 
results do not depend on the chosen gauge. We do not put $c=1$ in this and 
the following
appendices
to make more clear the order of terms with respect to the small 
parameter $1/c$. 

The ADM coordinate conditions, in linear approximation, read
\begin{eqnarray}\label{admgauge}
2\nabla_i g_{0i} - \nabla_0 g_{ii} = 0\;, \quad \quad
3\nabla_j g_{ij} - \nabla_i g_{jj} = 0\;,
\end{eqnarray}
where $\nabla_0=\partial/\partial t$ and $\nabla_i=\partial/\partial x^i$.
For comparison, the harmonic coordinate conditions, in linear approximation,
read:
\begin{eqnarray}\label{harmonic}
2\nabla_i g_{0i} - \nabla_0 g_{ii} = \nabla_0 g_{00}\;, \quad \quad
2\nabla_j g_{ij} - \nabla_i g_{jj} = - \nabla_i g_{00}\;.
\end{eqnarray}
The ADM gauge conditions (\ref{admgauge}) brings the space-space component of
metric to the form  
\begin{eqnarray}
\label{xxx}
g_{ij}&=& \delta_{ij} (1 + \frac{1}{3} h_{kk}) +
h^{TT}_{ij}\;,
\end{eqnarray}
where $h^{TT}_{ij}$ denotes the transverse-traceless part of $h_{ij}$.
Furthermore, in linear approximation, the Einstein field equations read
\begin{eqnarray}\label{eineq}
h_{00}&=& -\frac{8\pi}{c^4} \Delta^{-1}\left(T_{00} + T_{ii}\right)\;,\\
\nonumber\\\label{eineq2}
h_{0i}&=& -\frac{16\pi}{c^4} \Delta^{-1}\left(T_{0i} -
\frac{1}{4}\nabla_0\nabla_i \Delta^{-1}T_{00}\right)\;,\\
\nonumber\\\label{eineq3}
h_{kk}&=& - \frac{24\pi}{c^4} \Delta^{-1}T_{00}\;,\\
\nonumber\\\label{eineq4}
h^{TT}_{ij}&=& - \frac{16\pi}{c^4} P_{ijkl} \displaystyle{\Box^{-1}_{ret}}
T_{kl}\;,\\\nonumber
\end{eqnarray}
where the TT-projection operator $P_{ijkl}$, 
applied to symmetric tensors depending on both time and spatial coordinates, 
is given by
\begin{eqnarray}
P_{ijkl}&=&(\delta_{ik} - \Delta^{-1}\nabla_i \nabla_k)
              (\delta_{jl} - \Delta^{-1}\nabla_j \nabla_l) - 
\frac{1}{2} (\delta_{ij} - \Delta^{-1}\nabla_i \nabla_j)
              (\delta_{kl} - \Delta^{-1}\nabla_k \nabla_l)\;,
\end{eqnarray}
where $\Delta^{-1}$ denotes the Euclidean inverse Laplacian.

We now follow the calculation in the paper by 
Damour \& Esposito-Far\`ese \cite{23} using the ADM coordinate conditions.
The deflection of the light ray is given by
\begin{eqnarray}
\Delta {\it{l}}_{\mu} = {\it{l}}_{\mu}(+\infty) - {\it{l}}_{\mu}(-\infty)=
\frac{1}{2} \int^{+\infty}_{-\infty} d\tau {\it{l}}^{\alpha} {\it{l}}^{\beta}
\nabla_{\mu}h_{\alpha \beta}(\xi^{\lambda} + \tau
{\it{l}}^{\lambda}).\\\nonumber
\end{eqnarray}
In terms of the Fourier transform
\begin{eqnarray}\label{four}
{\hat h}_{\mu\nu}(k^{\lambda}) = \int d^4x h_{\mu\nu}(x^{\lambda}) ~
\mbox{e}^{-i k_{\alpha}x^{\alpha}},
\end{eqnarray}
where $k_{\alpha}x^{\alpha}=-k^0 x^0+{\bf k} \cdot {\bf{x}}$, the boldface
letters denote spatial components of vectors \cite{68}, and integration is
over all of space-time. Accounting for the formulae (\ref{four}) 
the equation for light deflection now reads 
\begin{eqnarray}
\Delta {\it{l}}_{\mu} =
i \pi \int \frac{d^4k}{(2\pi)^4} k_{\mu} ~ \mbox{e}^{ik_{\mu}\xi^{\mu}}
{\it{l}}^{\alpha} {\it{l}}^{\beta} {\hat h}_{\alpha \beta}(k^{\lambda})
\delta({\bf k} \cdot {\bf{l}}-k^0l^0),
\end{eqnarray}
where use has been made of the exact relationship
\begin{eqnarray}
\int_{-\infty}^{+\infty} d\tau ~ \mbox{exp}(i \tau k_{\alpha}l^{\alpha})
& =& 2\pi \delta \left({\bf k}\cdot {\bf l}-k^0l^0\right)\;.
\end{eqnarray}
In terms of the Fourier transformed energy-momentum tensor $\hat
T_{\alpha\beta}$, the Fourier transformed metric 
field reads
\begin{eqnarray}
\hat h_{00}&=& \frac{8\pi}{c^4} \frac{\hat T_{00} + \hat T_{ii}}{{\bf{k}}^2}\\
\nonumber\\
\hat h_{0i} &=& \frac{16\pi}{c^4} \left(\frac{\hat T_{0i}}{{\bf{k}}^2} -
\frac{1}{4} \frac{k_0 k_i}{{\bf{k}}^2} \frac{\hat T_{00}}{{\bf{k}}^2}\right)\\
\nonumber\\
\hat h_{kk} &=&  \frac{24\pi}{c^4}\frac{\hat T_{00}}{{\bf{k}}^2}\\
\nonumber\\
\hat h^{TT}_{ij} &=& \frac{16\pi}{c^4}\left[
(\delta_{il} - \frac{k_ik_l}{{\bf{k}}^2})(\delta_{jk} -
\frac{k_jk_k}{{\bf{k}}^2})
- \frac{1}{2}(\delta_{ij} - \frac{k_ik_j}{{\bf{k}}^2})(\delta_{kl} -
\frac{k_kk_l}{{\bf{k}}^2})\right]  
\frac{\hat T_{kl}}{{\bf{k}}^2 - (k^0 + i\epsilon)^2}\;,\\\nonumber
\end{eqnarray}
where $\epsilon$ is a positive infinitesimal number which shows
explicitly that we have used the retarded Green's function while solving the
Einstein equations for the component of the metric 
$h_{ij}$.
If we put ${\bf k}\cdot {\bf l}-k^0l^0=0$ and take into account the relations
\begin{eqnarray}
\hat T_{00}= \frac{k_ik_j}{k_0^2}\hat T_{ij}\;, \quad \quad
\hat T_{0i}= \frac{k_j}{k_0}\hat T_{ij}\;,
\end{eqnarray}
which follow from the macroscopic equations of motion for matter
$\nabla_{\nu}T^{\mu\nu} =0$, we find:
\begin{eqnarray}
{\it{l}^{\alpha}}{\it{l}^{\beta}}\hat h_{\alpha \beta}&=&
({\it{l}}^0)^2 (\frac{k_i}{k^0} - \frac{{\it{l}_i}}{{\it{l}}^0})
(\frac{k_j}{k^0} - \frac{{\it{l}_j}}{{\it{l}}^0})
\frac{\hat T_{ij}}{{\bf{k}}^2 - (k^0 + i\epsilon)^2}\;.
\end{eqnarray}
This expression is identical with that 
obtained by Damour \& Esposito-Far\`ese \cite{23} 
in the harmonic gauge. 
For this reason calculation of the total 
deflection angle gives
the same result in both harmonic and ADM gauges, reflecting the 
coordinate independence of the final result.
 
Defining $\alpha_{\mu} = \Delta{\it{l}}_{\mu}/{\it{l}}^0$, one gets for the
angle of total deflection
\begin{eqnarray}
\alpha_1& &= -\frac{4}{d^3}\left[
{\cal{I}}_{11}(t^*) - {\cal{I}}_{22}(t^*)\right],\\
\nonumber\\
\alpha_2 &=& \frac{8}{d^3}{\cal{I}}_{12}(t^*),\\
\nonumber\\
\alpha_3 &=&\alpha^0 =  -\frac{2}{d^2}
\left[\dot{\cal{I}}_{11}(t^*) - \dot{\cal{I}}_{22}(t^*)\right]\;.\\\nonumber
\end{eqnarray}
One can see that expressions for $\alpha_1$, $\alpha_2$ are the same as those 
obtained in section 7.1.2. The quantity $\alpha^0$ gives the gravitational shift 
in
frequency of the electromagnetic wave. It can be obtained from the expression
for gravitational time delay $\Delta_Q$ after its differentiation with respect
to time.

\section{Harmonic and ADM gauge conditions in the first
post-Minkowskian approximation}

In this Appendix we give other representations of the metric coefficients
(\ref{7}) - (\ref{9}). Using the ADM coordinate conditions (\ref{admgauge}) 
of Appendix A
the metric coefficients (\ref{7}) - (\ref{9}) can be cast into the ``canonical" ADM form
\begin{eqnarray}\label{a1a}
h_{00}^{adm}&=&\frac{2{\cal M}}{c^2r}+ \frac{{\cal{I}}_{ij}(t)}{c^2}
\nabla_i \nabla_j r^{-1},\\\nonumber\\\label{a2a}
h_{0i}^{adm}&=&-\frac{2}{c^3}\frac{\epsilon_{ipq}{\cal S}_p N_q}{r^2} +
\frac{2{\dot{\cal{I}}}_{ij}(t)}{c^3} \nabla_j r^{-1} -
\frac{{\dot{\cal{I}}}_{jk}(t)}{4c^3} \nabla_i \nabla_j \nabla_k r,\\\nonumber\\
h_{kk}^{adm}&=&3\;h_{00},\\\nonumber\\\label{a4a}
h_{ij}^{adm TT}&=&\frac{2\ddot{\cal{I}}_{ij}(t-r/c)}{c^4r}  \\
\nonumber\mbox{}&&+
\left(\delta_{ij} \nabla_k \nabla_l -
2 \delta_{il} \nabla_j \nabla_k - 2 \delta_{jl} \nabla_i \nabla_k\right)
\left[\frac{{\cal{I}}_{kl}(t-r/c)}{c^2r} -
\frac{{\cal{I}}_{kl}(t)}{c^2r}\right] + \\
\nonumber\\\nonumber\mbox{}&&
\nabla_i \nabla_j \nabla_k \nabla_l
\left[\frac{^{(-2)}{\cal{I}}_{kl}(t-r/c)}{r} -
\frac{^{(-2)}{\cal{I}}_{kl}(t)}{r} -
\frac{{\cal{I}}_{kl}(t)r}{2c^2}\right]\;,
\end{eqnarray}
where we have used a special symbolic notation for ``semihereditary 
functionals" \cite{69}
\begin{equation}\label{integr}
^{(-1)}{\cal{I}}_{ij}(t)\equiv\int_{-\infty}^t dv\;{\cal{I}}_{ij}(v)\;,
\quad\quad
^{(-2)}{\cal{I}}_{ij}(t)\equiv\int_{-\infty}^t dv\;^{(-1)}{\cal{I}}_{ij}(v)\;.
\\\nonumber
\end{equation}
The following equality holds:
$^{(-2)}\ddot{\cal{I}}_{ij}(t-r) = {\cal{I}}_{ij}(t-r)$.
We also notice that $\Delta ({\cal{I}}_{ij}(t-r/c)/r) =
\ddot{\cal{I}}_{ij}(t-r/c)/c^2r$ for $r \ne 0$. For this reason function 
$^{(-2)}\ddot{\cal{I}}_{ij}(t-r)$ is a solution of 
the homogeneous d'Alembert's
equation; that is, $\Box \left[^{(-2)}{\cal{I}}_{ij}(t-r)/r\right] =0$ 
for $r \ne 0$. 

We emphasize that the metric (\ref{a1a})-(\ref{a4a}) is an external 
solution of the
equations (\ref{eineq})-(\ref{eineq4}), outside 
the source of gravitational
waves. It matches smoothly to the internal solution, which is valid 
inside the
source, without additional coordinate transformations. It is
remarkable that outside the source the metric component (\ref{a4a}) may be 
represented as an
algebraic decomposition of the retarded and instantaneous functions of time
\begin{eqnarray}
\label{decomp}
h_{ij}^{adm TT}&=&h_{ij}^{TT}(t-\frac{r}{c},{\bf x})+\tilde{h}_{ij}(t,{\bf
x})\;,
\end{eqnarray}     
where $h_{ij}^{TT}(t-r/c,{\bf x})$ is shown below in (\ref{adm4}), and 
$\tilde{h}_{ij}(t,{\bf x})$ is the rest of the metric component 
$h_{ij}^{adm TT}$ which is actually a symmetrized gradient of a vector 
comprising of singular harmonic functions. For this reason function 
$\tilde{h}_{ij}(t,{\bf x})$ satisfies the condition 
$P_{ijkl}\;\tilde{h}_{kl}(t,{\bf x})\equiv 0$ and can be eliminated by an 
infinitesimal coordinate transformation. 
Making use of this  and 
without leaving the ADM coordinate conditions, we may construct the following
representation for the metric:
\begin{eqnarray}\label{adm1}
h_{00}&=&\frac{2{\cal M}}{c^2r},\\
\nonumber\\\label{adm2}
h_{0i}&=&-\frac{2}{c^3}\frac{\epsilon_{ipq}{\cal S}_p N_q}{r^2},\\
\nonumber\\\label{adm3}
h_{kk}&=&3\; h_{00},\\
\nonumber\\\label{adm4}
h^{TT}_{ij}&=& \frac{2\ddot{\cal{I}}_{ij}(t-r/c)}{c^4r} + \\
\nonumber\mbox{}&&
\left(\delta_{ij} \nabla_k \nabla_l -
2 \delta_{il} \nabla_j \nabla_k - 2 \delta_{jl} \nabla_i \nabla_k\right)
\frac{{\cal{I}}_{kl}(t-r/c)}{c^2r} +
\nabla_i \nabla_j \nabla_k \nabla_l\left[
\frac{^{(-2)}{\cal{I}}_{kl}(t-r/c)}{r}\right]\;.
\end{eqnarray}
This form of the metric is obtained from the expressions
(\ref{a1a})-(\ref{a4a}) by
applying the coordinate transformation \cite{70}
\begin{eqnarray}\label{c1}
w^0&=& \frac{1}{2}\nabla_k \nabla_l \left[
\frac{^{(-1)}{\cal{I}}_{kl}(t)}{cr}\right]\;,\\\nonumber\\\label{c2}
w^i&=&\frac{1}{2}\nabla_i \nabla_k \nabla_l\left[
\frac{^{(-2)}{\cal{I}}_{kl}(t)}{r}\right] -
2\nabla_k \left[\frac{{\cal{I}}_{ki}(t)}{c^2r}\right] +
\frac{1}{4}\nabla_i \nabla_k \nabla_l\left[
\frac{{\cal{I}}_{kl}(t)r}{c^2}\right]\;.
\end{eqnarray}
It is marvelous that this representation of metric 
also fulfils the harmonic coordinate conditions (\ref{harmonic}). 
This means
that outside the localized source of gravitational waves the class of ADM
coordinates overlaps with that of harmonic ones.  
The coordinate transformation from metric (\ref{7})-(\ref{9}) written in 
the pure harmonic coordinate system 
to the ADM-harmonic metric (\ref{adm1})-(\ref{adm4}) 
is:
\begin{eqnarray}\label{ttt}
w^0&=& \frac{1}{2}\nabla_k \nabla_l \left[
\frac{^{(-1)}{\cal{I}}_{kl}(t-r/c)}{cr}\right]\;,
\\\nonumber\\\label{kkk}
w^i&=&\frac{1}{2}  \nabla_i \nabla_k \nabla_l\left[
\frac{^{(-2)}{\cal{I}}_{kl}(t-r/c)}{r}\right] -
2\nabla_k \left[\frac{{\cal{I}}_{ki}(t-r/c)}{c^2r}\right]\;.
\end{eqnarray}
These gauge functions have been extensively used in the main body of the paper
for elaborating unique interpretation of observable effects. In contrast to the
expressions (B1)-(B4) the expressions (B6)-(B9) show terms which decay like
$1/r^4$ and $1/r^5$. These terms depend on time integrals of the quadrupole
moment ${\cal{I}}_{ij}(t-r/c)$ and may lead to the appearance of 
the ``kinematic resonance effect" discussed by Braginsky \& Grishchuk \cite{14}.
Another important remark is that the transformations
(\ref{ttt})-(\ref{kkk})
clearly show how to eliminate non-radiative terms from the metric, written down
in a harmonic gauge, including all terms with respect to any power of $1/r$.
Previously used transformations (see, for example, the textbook of Misner {\it
et al.} \cite{45}, paragraph 35) dealt only with terms of the first order in
$1/r$ and could not be applied for analysis of gravitational radiation in near
or intermediate zones of the localized source of gravitational waves.

\newpage
\begin{figure*}
\centerline{\psfig{figure=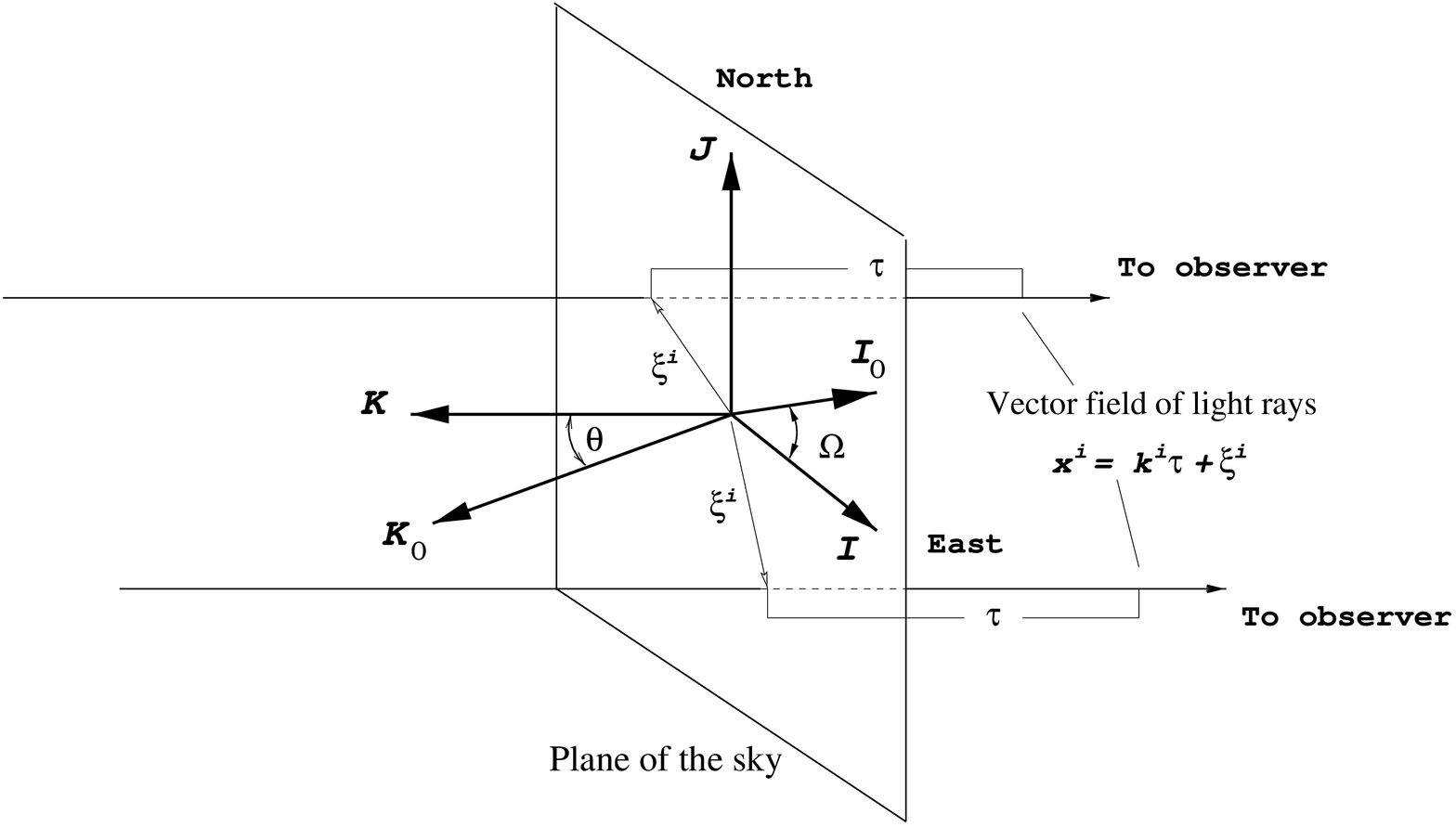,angle=270,height=12cm,width=18cm}}
\vspace{1 cm}
\caption{Astronomical coordinate 
system used for calculations. The origin of
the coordinate system is at the center-of-mass of the source of gravitational
waves. The bundle of light rays is defined by the vector field $k^i$. The 
vector $K^i=-k^i+O(c^{-2})$ is directed from observer towards the source 
of light. The
vector $K_0^i$ is directed from the observer towards the source of gravitational
waves. We use in the paper the equalities $K_0^i=-N^i=-x^i/r$, where $x^i$ are the
coordinates of the observer with respect to the source of gravitational waves, and
$r=|{\bf x}|$. The plane of the sky to the vector $K_0^i$ is not shown.}
\label{bundle}
\end{figure*}
\newpage
\begin{figure*}
\centerline{\psfig{figure=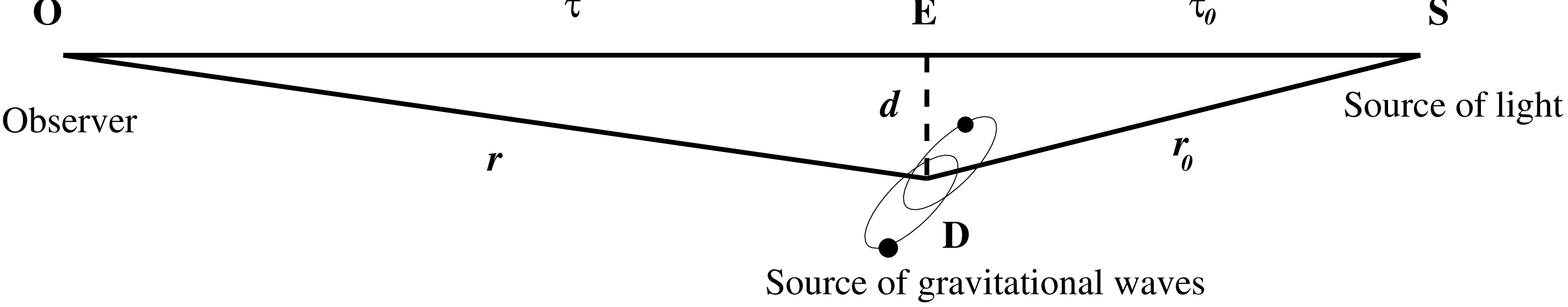,angle=270,height=3.5cm,width=16cm}}
\vspace{1 cm}
\caption{Relative configuration of observer (O), source of light (S), 
and a localized source of gravitational waves (D).
The source of gravitational waves deflects light 
rays which are
emitted at the moment $t_0$ at the point S and received at the moment $t$ at
the point O. The point E on the
line OS corresponds to the moment of the closest approach of light ray to the
deflector D. Distances are $OS=R$, $DO=r$, $DS=r_0$, the impact parameter $DE=d$, 
$OE=\tau>0$, $ES=\tau_0=\tau-R<0$. 
The impact parameter $d$ is small in comparison to all other distances.  }
\label{smallimp1}
\end{figure*}
\clearpage
\begin{figure*}
\centerline{\psfig{figure=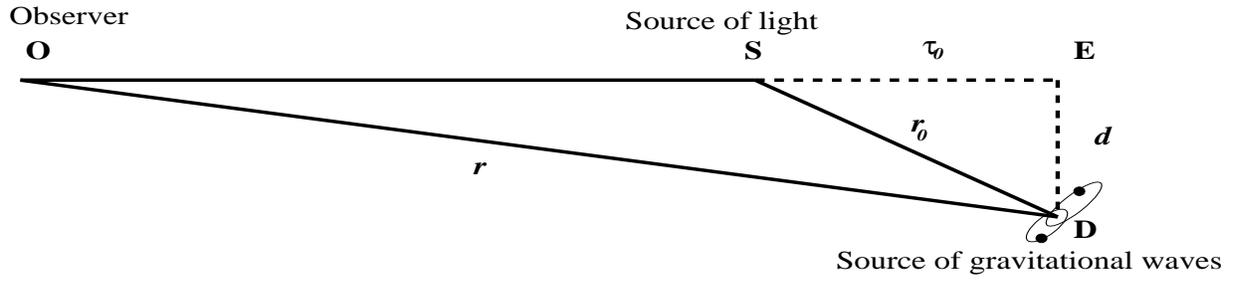,angle=270,height=3.5cm,width=16cm}}
\vspace{1 cm}
\caption{Relative configuration of observer (O), source of light (S), 
and a localized source of gravitational waves (D).
The source of gravitational waves deflects light 
rays which are
emitted at the moment $t_0$ at the point S and received at the moment $t$ at
the point O. The point E on the
line OS indicates the point of
minimal distance of the light ray trajectory extrapolated
backward to the
deflector D. Distances are $OS=R$, $DO=r$, $DS=r_0$, the impact parameter $DE=d$, 
$OE=\tau>0$, $ES=\tau_0=\tau-R>0$. 
The impact parameter $d$ is small in comparision to all other distances.  }
\label{smallimp2}
\end{figure*}
\clearpage
\begin{figure*}
\centerline{\psfig{figure=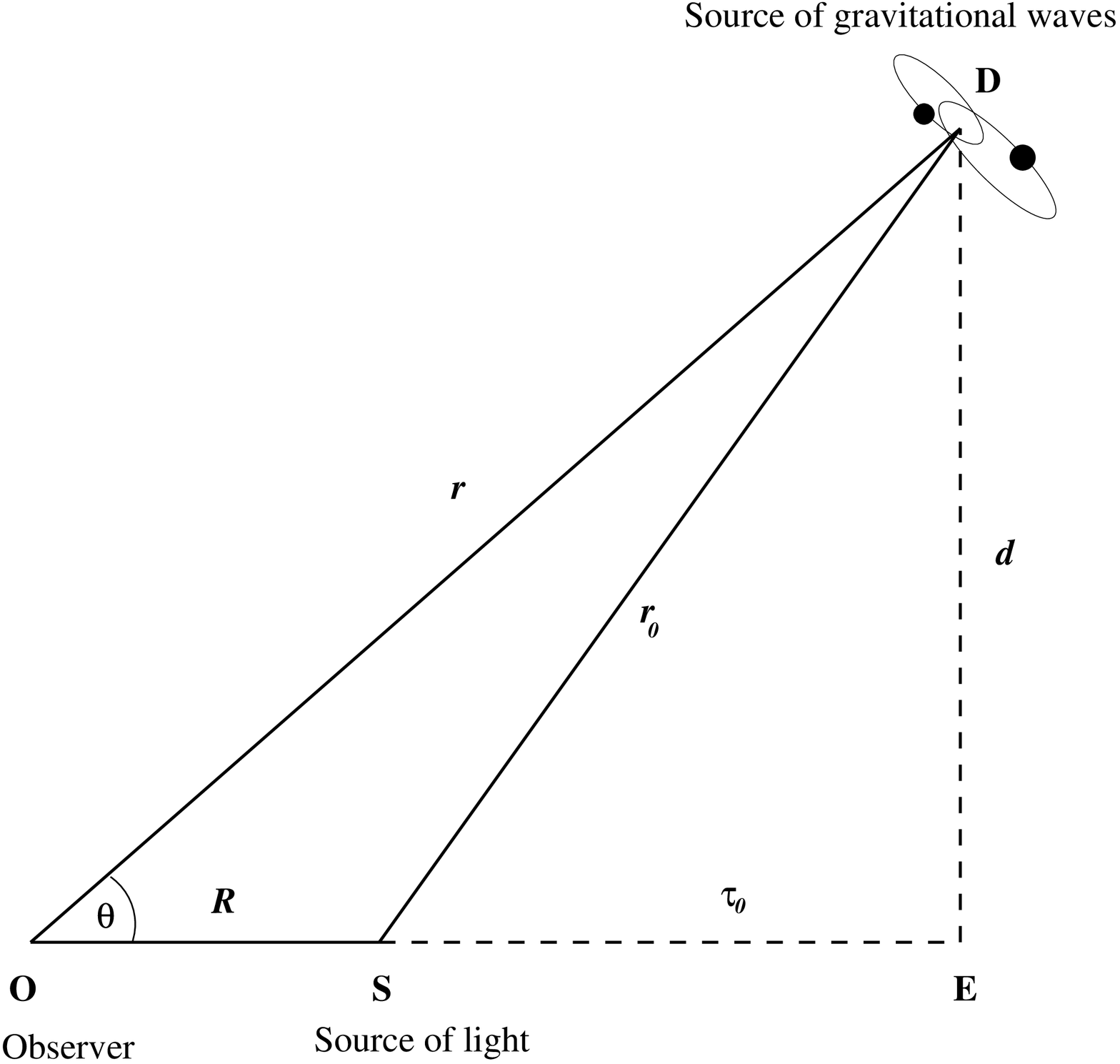,angle=270,height=8 cm,width=16cm}}
\vspace{1 cm}
\caption{Relative configuration of observer (O), source of light (S), 
and a localized source of gravitational waves (D).
The source of gravitational waves deflects light 
rays which are
emitted at the moment $t_0$ at the point S and received at the moment $t$ at
the point O. The point E on the
line OS corresponds to the moment of the closest approach of light ray to the
deflector D. Distances are $OS=R$,  $DO=r$, $DS=r_0$, the impact parameter $DE=d$, 
$OE=\tau=r\cos\theta$, $ES=\tau_0=\tau-R$. 
The distance $R$ is much smaller than both $r$ and $r_0$. 
The impact parameter $d$ is, in general, not small in 
comparision to all other distances. }
\label{largeimp}
\end{figure*}

\begin{thebibliography}{99}
\bibitem{36} Hereafter we put the universal 
gravitational constant $G=1$.
Repeated spatial indices in all subsequent equations mean summation from 1 to
3.
\bibitem{42} We draw attention to the minus sign  
in front of the  function $k^iw^0$ in equation (\ref{eqnm}). It
appears because of the  
equality $w_0=-w^0$, for we raise  and lower space-time 
indices using the Minkowski metric $\eta_{\alpha\beta}
={\rm diag}(-1,+1,+1,+1)$. Spatial 
indices are raised and lowered by means of
the Kronecker symbol $\delta_{ij}$. For this reason, there is no difference
between upper and lower spatial indices.
\bibitem{43} In fact, we must first calculate the 
derivative and then
integrate with respect to time. This eliminates the  
divergent part of the
integral (\ref{31}).
\bibitem{44} Damour \& Esposito-Far\`ese
\cite{23} have used in their calculations of light deflection the space-time 
Fourier transform of metric tensor. In our approach we use a Fourier
transform in the time domain only. This is simpler than the approach of 
Damour \& Esposito-Far\`ese \cite{23} and is complete enough for calculations
of observable effects.
\bibitem{45a} In higher approximations the outgoing gravitational radiation
leads to secular decrease in ${\cal M}$ and ${\cal S}$. We neglect this effect
in the present paper.
\bibitem{48} We  emphasize once again 
that vectors $K^i$ and $K^i_0$ defined in section 
$4.1$ determine different directions, as shown in 
Fig. \ref{bundle}.
\bibitem{52} The case of
observer moving with respect to the ADM-harmonic system with velocity $v^i$ may
be considered after completing the additional Lorentz transformation using 
the well-known procedure (see, for example, \cite{31}).
\bibitem{53} Terms 
proportional to $1/r_0$ do not appear in the expansion (\ref{quad}) while
terms of order $1/r_0^2$ do appear. The situation is different, however,
if the impact parameter is large (see section 7.2.3).
\bibitem{53b} Usually, it is assumed that the quadrupolar approximation
preassumes a slow-motion approximation \cite{23},
\cite{53a}. We would like to emphasize, however, that the slow-motion
approximation is not a direct consequence of
the quadrupolar approximation which only relies on the
assumption that the magnitudes of the octupole and other higher multipoles are
much smaller than that of the quadrupole. In principle, this does not prohibit 
velocities of matter to be close to velocity of light.
\bibitem{54} Those residual terms in this formulae
which depend only on the quadrupole moment decay as
either $1/r^2$, or $1/r^2_0$. It is remarkable that both terms of order $1/r$ 
and
those of $1/r_0$ do completely vanish because of cancellation of corresponding
terms.
\bibitem{59} A necessary 
condition for convergence  is the  assumption that the 
distance $R$ between source of light and observer is smaller than 
the wavelength 
$\lambda$ of gravitational waves emitted by the localized source, that is
$R\ll\lambda$. This condition holds, for instance, for ultra-long gravitational
waves of cosmological origin.
\bibitem{60} We recall that the origin of the coordinate system is at
the center of mass of the localized source emitting gravitational waves. This 
explains the appearance of the minus sign  in formula (\ref{vecxi}).
\bibitem{62} We can even 
apply our formalism to a general theoretical treatment of the  output of 
ground-based or space-borne
gravitational wave detectors such as LIGO or LISA. 
For instance, if we introduce the 
notation $\Delta R=c\Delta_Q$ for the variation of distance between the test
bodies of the detectors, then from formula (\ref{timedel}) and Taylor 
expansion of the quadrupole moment (\ref{texp}) one immediately obtains 
the well known result
\begin{eqnarray}
\nonumber
\frac{\Delta R}{R}&=&\frac{1}{2}k^ik^jh_{ij}^{TT}(t-r)+
O\left(\frac{R}{r}\right)+O\left(\frac{R}{\lambda}\right)\;,
\end{eqnarray}
where $\lambda$ is the wavelength of gravitational 
waves impinging on the detector. 
If necessary we can write down corrections to this formula up to any 
order. The case when $\lambda$ is comparable
with $R$ should be treated in a slightly different way and is a matter for a 
separate study. 
We emphasize that this can be done in the 
framework of our formalism as well.
\bibitem{63} These terms are proportional to the first time 
derivative of the metric perturbations $h_{ij}^{TT}(t-r)$.
\bibitem{64} Let us draw attention to the
fact that formula (\ref{direc}) is applicable to an arbitrary source of
gravitational waves, 
including the case of a plane gravitational wave, as has
been debated in the literature (see, for example, 
\cite{13}, \cite{15} and references therein). 
Unfortunately
there has been no agreement among the previous authors 
regarding explicit expression
for the light deflection angle. Formula (\ref{direc}) fill in this gap and
closes the debate.
\bibitem{65} Tsvetkov \& Tsyrulev \cite{66}
showed  the presence of a frequency $\Omega/2$ in the spectrum of gravitational
radiation emitted by a rapidly rotating fluid ball of a magnetized homogeneous
gravitating liquid which has an asymmetric form due to magnetic strains. We do 
not expect any such sources to be prominent at low
frequencies.
\end{thebibliography}
\end{document}